\documentclass[aps,prc,letterpaper,11pt,twoside,tightenlines,nofootinbib,showpacs
,preprint,superscriptaddress]{revtex4-1}
\usepackage{epsfig,float}
\usepackage{amsmath}
\begin{document}

\newcommand{\Ima}{\textrm{Im}}
\newcommand{\Rea}{\textrm{Re}}
\newcommand{\mev}{\textrm{ MeV}}
\newcommand{\gev}{\textrm{ GeV}}
\newcommand{\rb}[1]{\raisebox{2.2ex}[0pt]{#1}}

\title{Baryon states with open beauty in the extended local hidden gauge approach.}
%\author{Weihong Liang$^{1,2}$, Chu Wen Xiao$^1$ and E. Oset$^1$}
%\affiliation{$^1$ Departamento de F\'{\i}sica Te\'orica and IFIC, Centro Mixto Universidad de Valencia-CSIC,
%Institutos de Investigaci\'on de Paterna, Aptdo. 22085, 46071 Valencia,Spain}
%\affiliation{$^2$ Department of Physics, Guangxi Normal University, Guilin, 541004, P. R. China}

\author{W. H. Liang}
\email{liangwh@gxnu.edu.cn}
\affiliation{Department of Physics, Guangxi Normal University, Guilin, 541004, P. R. China}
\affiliation{
Departamento de F\'{\i}sica Te\'orica and IFIC, Centro Mixto Universidad \\de Valencia-CSIC, Institutos de Investigaci\'on de Paterna, Apartado 22085, 46071 Valencia, Spain}

\author{C. W. Xiao}
\author{E. Oset}
\affiliation{
Departamento de F\'{\i}sica Te\'orica and IFIC, Centro Mixto Universidad \\de Valencia-CSIC, Institutos de Investigaci\'on de Paterna, Apartado 22085, 46071 Valencia, Spain}

\date{\today}

\begin{abstract}

In this paper we examine the interaction of $\bar B N$, $\bar B \Delta$, $\bar B^* N$ and $\bar B^* \Delta$ states, together with their coupled channels, using a mapping from the light meson sector. The assumption that the heavy quarks act as spectators at the quark level automatically leads us to the results of the heavy quark spin symmetry for pion exchange and reproduces the results of the Weinberg Tomozawa term, coming from light vector exchanges in the extended local hidden gauge approach. With this dynamics we look for states dynamically generated from the interaction and find two states with nearly zero width, which we associate to the $\Lambda_b(5912)$ and $\Lambda_b(5920)$ states. The states couple mostly to $\bar B^* N$, which are degenerate with the Weinberg Tomozawa interaction. The difference of masses between these two states, with $J=1/2,\ 3/2$ respectively, is due to pion exchange connecting these states to intermediate $\bar B N$ states. In addition to these two $\Lambda_b$ states, we find three more states with $I=0$, one of them nearly degenerate in two states of $J=1/2,\ 3/2$. Furthermore we also find eight more states in $I=1$, two of them degenerate in $J=1/2, 3/2$, and other two degenerate in $J=1/2,\ 3/2,\ 5/2$.

\end{abstract}


\maketitle

\section{Introduction}

Hadron Physics in the charm and beauty sectors is booming, with mounting activity in experiments BABAR, CLEO, BELLE, BES, LHC$_b$, CDF \cite{Ali:2011vy,Gersabeck:2012rp,Olsen:2012zz,bes,DeSimone:2012kxa,Palni:2012zxa} and theory \cite{Brambilla:2010cs}. One of the issues that has attracted much attention is the finding of hadronic states which cannot be interpreted in the conventional picture of $q \bar q$ for mesons and $qqq$ for baryons. Multiquark states, hybrids or hadronic molecules have been suggested in several works \cite{Kolomeitsev:2003ac,Maiani:2005pe,Guo:2006fu,Swanson:2005tq,Rosner:2006sv,Gamermann:2006nm, Nielsen:2009uh,Branz:2009yt,Klempt:2009pi,Ortega:2012rs,HidalgoDuque:2012pq}. The molecular picture stands on firm grounds once the use of chiral unitary theory in the light quark sector, or its extension through the local hidden gauge approach, has shown that many mesonic and baryonic resonances are dynamically generated from the interaction of more elementary hadron components \cite{review,revhidden}. Concerning baryonic resonances with charm or hidden charm, work on molecules has been done in Refs. \cite{Hofmann:2005sw,Mizutani:2006vq,Tolos:2007vh,Wu:2010jy,Wu:2010vk,Romanets:2012ce,Garcia-Recio:2013gaa,Romanets:2012hm,xiaojuan}, while in the beauty sector, baryon states with beauty or hidden beauty have also been studied in Refs. \cite{wuzou,juanloren,xiaooset,david}.

On the experimental side, $\Lambda_b$ excited states have been reported by the LHC$_b$ collaboration in Ref. \cite{Aaij:2012da}. Two states, $\Lambda_b(5912)$ and $\Lambda_b(5920)$ are found in the experiment, with widths smaller than 0.66 MeV in both cases. Although no direct spin and parity have been determined, the states are interpreted as orbitally excited states of the ground state of the $\Lambda_b(5619)$ . One of the states, the $\Lambda_b(5920)$, has been confirmed by the CDF collaboration in Refs. \cite{Aaltonen:2013tta,Palni:2013bza}.
The association to the orbitally excited states of the $\Lambda_b(5619)$ seems most natural since predictions of quark models had been done for these states, as the orbitally excited $\Lambda_b$ states with $L=1$ and $J^P=1/2^-,3/2^-$ \cite{Capstick:1986bm,Garcilazo:2007eh}. Compared to the observed results, the $\Lambda_b$ masses, including that of the ground state, are only off by about 30-35 MeV.

  The closest work in spirit to the present one is that of Ref. \cite{juanloren} where these states are dynamically generated from the interaction of mesons and baryons. In  Ref. \cite{juanloren} the Heavy Quark Spin Symmetry (HQSS) is used as an underlying symmetry. According to it, the $B$, $B^*$ states are degenerate in the heavy quark limit, as well as the $J^P=1/2^+, 3/2^+$ baryon states, which are then considered together in a coupled channels approach. An extrapolation of the Weinberg Tomozawa interaction in the light sector is then used \cite{juanloren,Romanets:2012hm}, with elements of an SU(6) spin-isospin symmetry \cite{GarciaRecio:2008dp}. With suitable choices of the renormalization scheme for the loops, good agreement with the masses of the newly found  $\Lambda_b$ states is obtained. Our scheme takes advantage of the study done in Refs.
\cite{xiaojuan,xiaooset,xiaoaltug}. In those works it was found that the use of the extended local hidden gauge approach to the heavy quark sector fully respects the HQSS, but it provided a dynamics different from the one of Ref. \cite{juanloren}. In particular, the connection between $B$ and $B^*$ states (or baryon states with $J^P=1/2^+, 3/2^+$) requires pion exchange, or anomalous terms, which are found subleading in the large heavy quark mass counting, and numerically small. Similar conclusions are also found in Ref. \cite{Nieves:2011zz}. In some works \cite{slzhu}, pion exchange is found relevant compared to vector meson exchange (the dominant terms in the local hidden gauge approach), but as discussed in Ref. \cite{xiaoaltug}, this is in part due to the use of a type of form factor for vector mesons, not present when the equivalent chiral amplitudes are constructed, that suppresses the vector exchange.

With the dynamics of the extended local hidden gauge approach, the number of coupled channels is small for each quantum number and we can get a good feeling of the basic building blocks in the dynamically generated states that appear. As we shall see, the two $\Lambda_b$ excited states are generated with masses close to the experimental ones, but more remarkable, the difference in the masses of the two excited states agrees well with experiment. The widths obtained are zero within our basis of coupled channels, quite in agreement with the widths smaller than 0.66 MeV found in the experiment.  We also make predictions in other isospin and spin sectors. On the formal aspects we show how the dominant terms correspond to having a heavy quark as spectator and present an easy way to implement the results of the heavy quark spin-flavour symmetry from the impulse approximation at the quark level.

\section{Formalism}

\begin{figure}[tb]
\epsfig{file=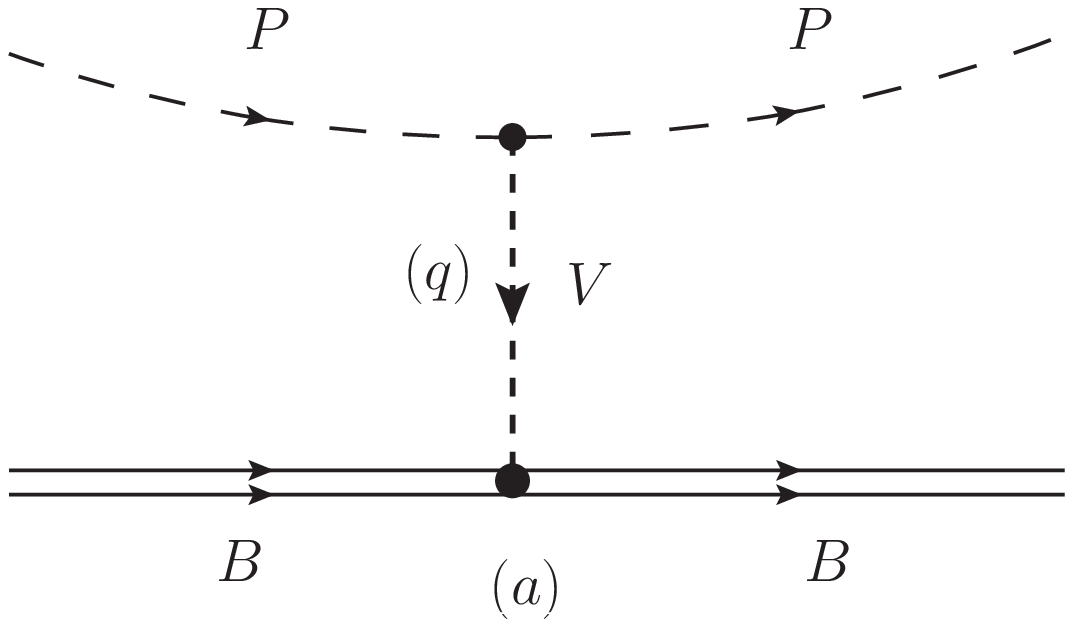, width=6.5cm}\hspace{1.1cm} \epsfig{file=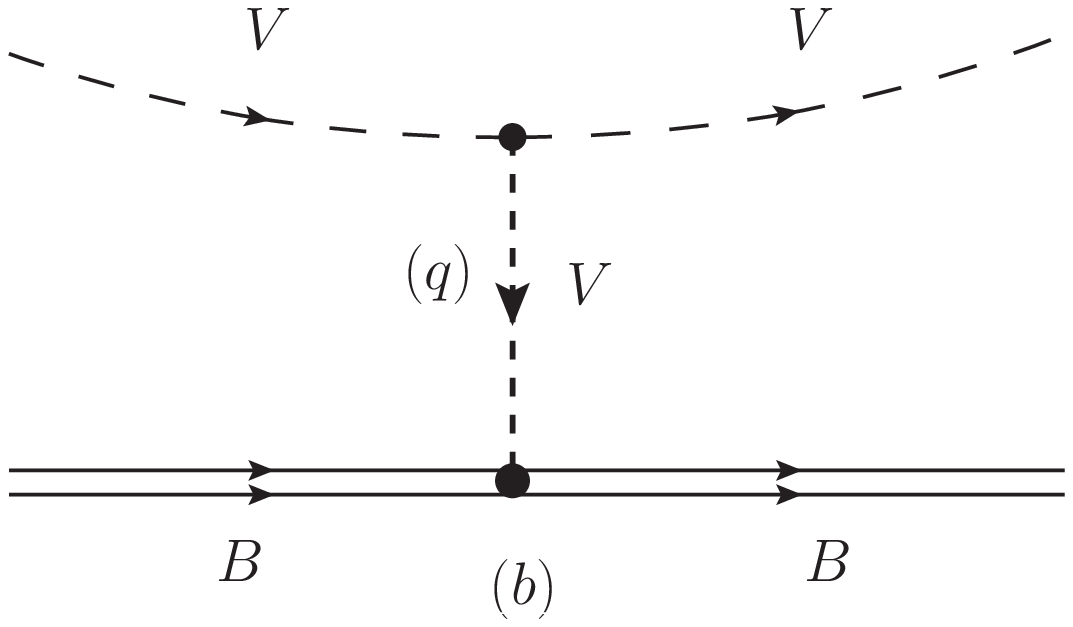, width=6.5cm}
\caption{Diagrammatic representation of the pseudoscalar baryon interaction (a) and vector baryon interaction (b).}\label{fig:f1}
\end{figure}

We will look at the states $\pi \Sigma_b$, $\pi \Lambda_b$, $\eta \Lambda_b$, $\eta \Sigma_b$, $\bar{B} N$ which can couple to $I=0, ~1$ which we will investigate. Similarly, we shall look at $\bar{B}^* N$ and $\pi \Sigma_b^*$, $\eta \Sigma_b^*$, $\bar{B} \Delta$, $\bar{B}^* \Delta$, with $\Delta \equiv \Delta (1232)$ and $\Sigma_b^* = \Sigma_b^* (5829)$, belonging to a decuplet of $3/2^+$ states. In the local hidden gauge approach in SU(3) \cite{hidden1,hidden2,hidden4} the meson baryon interaction proceeds via the exchange of vector mesons as depicted in Fig. \ref{fig:f1}.
As discussed in Ref. \cite{xiaojuan}, when we exchange a light vector meson in diagram (a), (b) of Fig. \ref{fig:f1}, the heavy quarks of the meson or the baryon are spectators and hence the interaction does not depend on their spin nor its flavour. From the technical point of view the interaction of the diagrams of Fig. \ref{fig:f1} can be obtained using SU(3) symmetry considering $u, ~d, ~b$ quarks, since we do not consider states with strangeness or hidden strangeness. Thus, all matrix elements of the interaction are formally identical (except for the mass or energy dependence) to those found for the interaction of the analogous states $\pi \Sigma$, $\pi \Lambda$, $\eta \Lambda$, $\eta \Sigma$, $\bar{K} N$, $\bar{K}^* N$, $\pi \Sigma^*$, $\eta \Sigma^*$, $\bar{K} \Delta$, $\bar{K}^* \Delta$. This interaction has been studied in Ref. \cite{angels} and Ref. \cite{sarkar}.

The transition potential from channel $i$ to channel $j$ is given by \cite{bennhold}
\begin{equation}
V = -C_{ij} \frac{1}{4f^2} (2 \sqrt{s} - M_{B_i} - M_{B_j}) \sqrt{\frac{M_{B_i} + E_i}{2 M_{B_i}}} \sqrt{\frac{M_{B_j} + E_j}{2 M_{B_j}}},
\label{eq:vij}
\end{equation}
with $f$ the pion decay constant, $M_{B_i}, ~E_i$ ($M_{B_j}, ~E_j$) the mass, energy of baryon of $i$ ($j$) channel. We take $f=f_\pi = 93\mev$ since we exchange light vector mesons. The $C_{ij}$ coefficients are evaluated in Refs. \cite{angels,sarkar} and we quote them below.

For pseudoscalar mesons and $1/2^+$ baryons we have the coupled channels $\bar{B} N$, $\pi \Sigma_b$, $\eta \Lambda_b$ in $I=0$ and the $C_{ij}$ coefficients are given in Table \ref{tab:vij0}.
\begin{table}[H]
     \renewcommand{\arraystretch}{1.5}
     \setlength{\tabcolsep}{0.4cm}
\centering
\begin{tabular}{c|ccc}
$C_{ij}$ & $\bar{B} N$ &  $\pi \Sigma_b$ & $\eta \Lambda_b$  \\
\hline
$\bar{B} N$ &  3 & $-\sqrt{\frac{3}{2}}$ & $\frac{3}{\sqrt{2}}$   \\
$\pi \Sigma_b$ &   & 4  & 0   \\
$\eta \Lambda_b$ &  &  & 0
\end{tabular}
\caption{$C_{ij}$ coefficients for $I=0$ and $J^P=1/2^-$.}
\label{tab:vij0}
\end{table}
In $I=1$ we have the channels $\bar{B} N$, $\pi \Sigma_b$, $\pi \Lambda_b$, $\eta \Sigma_b$ and the $C_{ij}$ coefficients are given  in Table \ref{tab:vij1}.
\begin{table}[H]
     \renewcommand{\arraystretch}{1.5}
     \setlength{\tabcolsep}{0.4cm}
\centering
\begin{tabular}{c|cccc}
$C_{ij}$ & $\bar{B} N$ &  $\pi \Sigma_b$ &  $\pi \Lambda_b$ & $\eta \Sigma_b$  \\
\hline
$\bar{B} N$ &  1 & $-1$ & $-\sqrt{\frac{3}{2}}$ & $-\sqrt{\frac{3}{2}}$   \\
$\pi \Sigma_b$ &   & 2  & 0  & 0 \\
$\pi \Lambda_b$ &  &  & 0 & 0  \\
$\eta \Sigma_b$&  &  & & 0
\end{tabular}
\caption{$C_{ij}$ coefficients for $I=1$ and $J^P=1/2^-$.}
\label{tab:vij1}
\end{table}
As one can see, the interaction in $I=0$ is stronger than that in $I=1$ and we have more chances to bind states in $I=0$.

As discussed in Ref. \cite{xiaojuan}, the mixing of states containing baryons of the octet (in $u, ~d, ~b$) like $\Sigma_b$ and of the decuplet $\Sigma_b^*$ require pion exchange for their mixing and this is strongly suppressed in the heavy quarks sector, hence, we neglect the mixing in a first step, but we shall come back to it in Section \ref{sec:mixing}. Then, if we consider a pseudoscalar meson and a baryon of the decuplet, we have the results for $C_{ij}$ given in Tables \ref{tab:vij0j3} and \ref{tab:vij1j3} \cite{sarkar}. We note that the strength of the $\bar{B} \Delta \to \bar{B} \Delta$ coefficient is four times bigger than for $\bar{B} N \to \bar{B} N$ and thus, we expect larger bindings in this case.
\begin{table}[H]
     \renewcommand{\arraystretch}{1.5}
     \setlength{\tabcolsep}{0.4cm}
\centering
\begin{tabular}{c|c}
$C_{ij}$ &  $\pi \Sigma_b^*$   \\
\hline
$\pi \Sigma_b^*$ &  4
\end{tabular}
\caption{$C_{ij}$ coefficients for $I=0$ and $J^P=3/2^-$.}
\label{tab:vij0j3}
\end{table}

\begin{table}[H]
     \renewcommand{\arraystretch}{1.5}
     \setlength{\tabcolsep}{0.4cm}
\centering
\begin{tabular}{c|ccc}
$C_{ij}$ & $\bar{B} \Delta$  &  $\pi \Sigma_b^*$  &  $\eta \Sigma_b^*$  \\
\hline
$\bar{B} \Delta$  &  4  & 1 & $\sqrt{6}$   \\
$\pi \Sigma_b^*$  &   & 2  & 0   \\
$\eta \Sigma_b^*$  &  &  & 0
\end{tabular}
\caption{$C_{ij}$ coefficients for $I=1$ and $J^P=3/2^-$.}
\label{tab:vij1j3}
\end{table}

The interaction $\bar{B} \Delta$ and coupled channels with $I=2$ is repulsive and we do not consider it.

In coupled channels we will use the Bethe-Salpeter equation
\begin{equation}
T = [1 - V \, G]^{-1}\, V,
\label{eq:Bethe}
\end{equation}
with $G$ the diagonal loop function for the propagating intermediate meson baryon channels. In Ref. \cite{xiaooset} we warned about potential dangers of using the dimensional regularization for the $G$ functions (see also Ref. \cite{wuzou}) since for values of the energy below threshold $G$ can soon become positive and then one can be misled to obtain bound states with a positive (repulsive) potential when $1 - VG =0$ (see Eq. \eqref{eq:Bethe} in one channel). For this reason we also use here the cut off regularization for $G$ given by
\begin{equation}
G(s) = \int_0^{q_{max}} \frac{d^{3}\vec{q}}{(2\pi)^{3}}\frac{\omega_P+\omega_B}{2\omega_P\omega_B}\,\frac{2M_{B}}{P^{0\,2}-(\omega_P+\omega_B)^2+i\varepsilon},
\label{eq:Gco}
\end{equation}
where  $\omega_P = \sqrt{\vec{q}\,^2+m_P^2},~\omega_B = \sqrt{\vec{q}\,^2+M_B^2}$, and $q_{max}$ is the cut-off of the three-momentum. However, in Ref. \cite{xiaooset}, we also took into account the form factor from vector meson exchange, by introducing
\begin{equation}
f(\vec{q}\,)=\frac{m_V^2}{\vec{q}\,^2+m_V^2}, \label{eq:ff}
\end{equation}
in which case we would have to replace Eq. \eqref{eq:Gco} by
\begin{equation}
G(s) = \int \frac{d^{3}\vec{q}}{(2\pi)^{3}} f(\vec{q}\,) \frac{\omega_P+\omega_B}{2\,\omega_P\,\omega_B}\,\frac{2M_{B}}{P^{0\,2}-(\omega_P+\omega_B)^2+i\varepsilon},
\label{eq:Gco2}
\end{equation}
putting the extra $f(\vec{q}\,)$ factor. We would like to make a comment here since in Ref. \cite{xiaooset} we put $f^2(\vec{q}\,)$. From the practical point of view, the differences between the two choices are smaller than uncertainties we will accept from other sources. From the theoretical point of view, while the first loop implicit in Eq. \eqref{eq:Bethe}, $VGV$, contains $f^2(\vec{q}\,)$, the terms in the series go as $VGV$, $VGVGV$, $VGVGVGV \cdots$ and the ratio of one term to the other is $GV$. Hence it is more appropriate to take just the one form factor of the potential $V$ and include it in the $G$ function when integrating over $\vec{q}$.

Since the $G$ function in Eq. \eqref{eq:Gco} is logarithmically divergent, the inclusion of $f(\vec{q}\,)$ in Eq. \eqref{eq:Gco2} makes it already convergent. Yet we will put an extra cut off $q_{max}$ that will serve to fine tune our $T$ matrix and the binding of the states. We shall fine tune $q_{max}$ in the integral of Eq. \eqref{eq:Gco} and we shall need values $q_{max}$ smaller than $M_V \approx 780 \mev$. Hence, from the practical point of view we can  even neglect the factor $f(\vec{q}\,)$ and effectively include its effects with the use of a suited value of $q_{max}$.

\begin{figure}[tb]
\epsfig{file=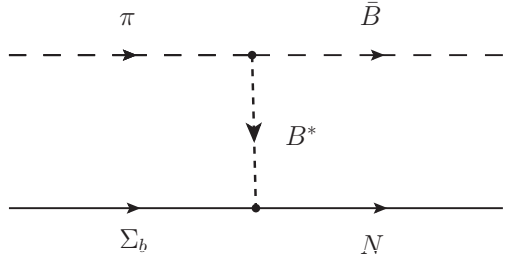, width=7cm}
\caption{Transition potential from $\pi \Sigma_b \to \bar{B} N$.}
\label{fig:Bex}
\end{figure}

Before closing this sector we must say two words concerning the transition $\pi \Sigma_b \to \bar{B} N$. This is depicted in Fig. \ref{fig:Bex}.
and it is mediated by $B^*$ exchange in the extended local hidden gauge approach. In the strict large heavy quark mass counting this term would be neglected because it involves the exchange of a heavy vector $B^*$ and its propagator would render this term negligible. However, although suppressed, it is not so much as one would expect. Indeed the propagator will be
\begin{equation}
D_{B^*} = \frac{1}{p^2_{B^*} - m^2_{B^*}} \equiv \frac{1}{(p^0_{\pi} - p^0_{\bar{B}})^2 - (\vec{p}_{\pi}-\vec{p}_{\bar{B}})^2 - m^2_{B^*}}.
\end{equation}
By contrary, in a diagonal transition $\bar{B} N \to \bar{B} N$ mediated by $\rho$ exchange for instance we would have
\begin{equation}
D_\rho \approx \frac{1}{m_V^2}.
\end{equation}
Close to $\bar{B} N$ threshold the ratio is
\begin{equation}
\frac{D_{B^*}}{D_\rho} \simeq \frac{m_V^2}{(p^0_{\pi} - p^0_{\bar{B}})^2 - \vec{p}_{\pi}^2 - m^2_{B^*}} \simeq \frac{1}{4}.
\end{equation}
Since the non diagonal terms have a smaller importance in the process than the diagonal ones of the heavy mesons, we simply account for these transitions multiplying by $1/4$ the results obtained from Eq. \eqref{eq:vij} and the Tables.

\section{Results for $I=0$}

We first choose the single channel $\pi \Sigma_b^*$ in $I=0$ and look for the binding energy. The state with $L=0$ has $J=3/2$. First we find that with the normal potential and a wide range of cutoffs (up to $3000\mev$) we do not find a bound state. We must look at the reason for this in the fact that the potential is indeed weak. This is so because the potential in Eq. \eqref{eq:vij} is a relativistic form of $k^0 + k'^0$ (the sum of the incoming and outgoing pion energies). The small mass of the $\pi$ makes its energy small close to threshold and this potential is subleading with respect to the one of $\bar{B} N$ where the energies now are those of the $\bar{B}$.

Next we try to see if increasing the potential by a factor $1.5$ or $2$ and varying the cut off we can obtain a reasonable binding. The results are chosen in Table \ref{tab:polepSs0}.
\begin{table}[H]
     \renewcommand{\arraystretch}{1.5}
     \setlength{\tabcolsep}{0.4cm}
\centering
\begin{tabular}{|c|c|c|c|c|c|c|c|c|}
\hline
$q_{max}$ & 800  &  1000  &  1200 & 1400 & 1600 & 1800 & 2000 & 3000  \\
\hline
$1.5 \,V$  &  5971  & 5965 & 5961  & 5956  & 5953  & 5950  & 5948 & 5942  \\
\hline
$2 \,V$  & 5955 & 5947 & 5940 & 5935 & 5932 & 5929 & 5927 & 5920 \\
\hline
\end{tabular}
\caption{Energies for $\pi \Sigma_b^*$ only channel as a function of $V$ and $q_{max}$. (unit: MeV)}
\label{tab:polepSs0}
\end{table}
As we can see, we have to increase the potential by a factor of two and go to very large cutoffs to obtain the desired value of the binding of the $\Lambda_b (5920)$. We might think that and increase by about a factor 1.5 of the potential could be accepted by recalling that such changes appear in models like Dyson Schwinger approach \cite{ElBennich:2011py} (see also Ref. \cite{xiaojuan}). Indeed, with respect to the coupling we would be using here, the $D \rho D$ coupling used in  Ref. \cite{ElBennich:2011py}, or in Ref. \cite{Can:2012tx} obtained with sum rules, is about a factor 1.5 bigger. However, in the same work of Ref. \cite{ElBennich:2011py}, the coupling is accompanied by a form factor which would be equivalent to a cut off $q_{max}$ of about 700 MeV. Hence, we cannot invoke simultaneously an increase of the potential by a factor 2 and a $q_{max}$ of 3000 MeV, and the only conclusion is that the $\pi \Sigma_b^*$ channel by itself cannot account for the $\Lambda_b (5920)$ state.

Next we repeat the same exercise with the single channel $\bar{B} N$ and show the results in Table \ref{tab:poleBbN0}.
\begin{table}[H]
     \renewcommand{\arraystretch}{1.5}
     \setlength{\tabcolsep}{0.4cm}
\centering
\begin{tabular}{|c|c|c|c|c|c|c|c|c|}
\hline
$q_{max}$ & 700 & 800  &  1000  &  1200 & 1400 & 1600 & 1800 & 2000  \\
\hline
$1 \,V$  & 6074 &  6026  & 5933 & 5851  & 5782  & 5725  & 5678  & 5639 \\
\hline
$1.5 \,V$  & 5967 &  5896  & 5766 & 5658  & 5572  & 5504  & 5450  & 5406 \\
\hline
$2 \,V$  & 5871 & 5784 & 5630 & 5509 & 5415 & 5343 & 5287 & 5243 \\
\hline
\end{tabular}
\caption{Energies for $\bar{B} N$ only channel as a function of $V$ and $q_{max}$. (unit: MeV)}
\label{tab:poleBbN0}
\end{table}
What we see in this table is that the binding grows spectacularly (and unrealistically) for bigger $V$ and $q_{max}$. Obviously the large value of the potential, as we mentioned above, is responsible for this. At this point we should mention that in the study of the $\bar{K} N$ system in coupled channels a cut off of $630\mev$ was used in Ref. \cite{angels}. In the study of the pseudoscalar mesons with the decuplet of baryons \cite{sarkar} a value of $q_{max} = 700\mev$ was used, while in Ref. \cite{wuzou} in the study of baryons with hidden beauty a value of $q_{max} = 800\mev$ was used. We can also see in Table \ref{tab:polepSs0}, that changes in $V$ can be accommodated by a change in $q_{max}$. In what follows we shall then use the potential that we get in the approach, without the extra multiplicative factor, but play with values of $q_{max}$ around $700 \mev - 850 \mev$, in the range of values used in previous works.

In most of the cases, we get energies where all the coupled channels are closed and, hence, the width is zero. When there are open channels we look for poles in the second Riemann sheet, which is obtained by changing the $G$ function as \cite{Roca:2005nm}
\begin{equation}
G_l^{II} (\sqrt{s}) = G_l^{I} (\sqrt{s}) + i\frac{q_l}{4 \pi \sqrt{s}}, \label{eq:rs2}
\end{equation}
where $q_l$ is the on shell momentum of the particles in the open channel, and $G_l^{I} (\sqrt{s})$ is given by Eq. \eqref{eq:Gco}.

As an example we show next the results without form factor of Eq. \eqref{eq:ff} for $\bar{B} N$, just changing $q_{max}$, as shown in Table\ref{tab:poleBbN0ng}.
\begin{table}[H]
     \renewcommand{\arraystretch}{1.5}
     \setlength{\tabcolsep}{0.5cm}
\centering
\begin{tabular}{|c|c|c|c|c|}
\hline
$q_{max}$ & 700  &  750  &  800 & 850  \\
\hline
$V$  &  5987.5  & 5941.6 & 5893.5  & 5843.7  \\
\hline
\end{tabular}
\caption{Energies for a state of $\bar{B} N$ in $I=0$ as a function of $q_{max}$. (unit: MeV)}
\label{tab:poleBbN0ng}
\end{table}

Next we introduce the coupled channels that couple to $\bar{B} N$ in $I=0$ (see Table \ref{tab:vij0}). The results that we obtained for the energy are shown in Table \ref{tab:polepCoup0}.
\begin{table}[H]
     \renewcommand{\arraystretch}{1.5}
     \setlength{\tabcolsep}{0.3cm}
\centering
\begin{tabular}{|c|c|c|c|c|}
\hline
$q_{max}$ & 700  &  750  &  800 & 850  \\
\hline
  &  5935.3  & 5897.3 & 5851.4  & 5802.0  \\
\cline{2-5} \rb{$V$} &  $6005.8+i23.8$  & $5988.9+i26.4$ & $5976.9+i24.4$  & $5968.0+i20.5$ \\
\hline
\end{tabular}
\caption{Energies for a state in coupled channels $\bar{B} N$, $\pi \Sigma_b$, $\eta \Lambda_b$ in $I=0$ as a function of $q_{max}$. (unit: MeV)}
\label{tab:polepCoup0}
\end{table}

The results are interesting. We see now that we get two states rather than one. In order to get a feeling of the meaning of the states we calculate the coupling of those states to the different coupled channels. We show the results in Table \ref{tab:cou10} for $q_{max}=800\mev$.
\begin{table}[ht]
     \renewcommand{\arraystretch}{1.2}
     \setlength{\tabcolsep}{0.3cm}
\centering
\begin{tabular}{cccc}
\hline\hline
$5851.4+i0$ & $\bar{B} N$ & $\pi \Sigma_b$ & $\eta \Lambda_b$  \\
\hline
$g_i$ & $16.20$ & $0.96$ & $1.47$  \\
$g_i\,G_i^{II}$ & $-20.55$ & $-16.23$ & $-14.31$  \\
\hline
$5976.9+i24.4$ & $\bar{B} N$ & $\pi \Sigma_b$ & $\eta \Lambda_b$  \\
\hline
$g_i$ & $5.88-i0.24$ & $1.52+i0.75$ & $0.75+i0.02$  \\
$g_i\,G_i^{II}$ & $-9.60-i0.16 \,$ & $\, -53.13-i12.34 \,$ & $\, -9.33-i0.85$  \\
\hline
\end{tabular}
\caption{The coupling constants to various channels for certain poles in the $J=1/2,~I=0$ sector.} \label{tab:cou10}
\end{table}
We show the values of the couplings ($g_i^2$ is the residue of the matrix element $T_{ii}$ at the pole) and of $g_i G_i^{II}$, which, according to \cite{gamerjuan}, provides the wave function of the origin in coordinate space, the magnitude that shows the relevance of the channel in the short range strong interactions. It is interesting to see that there has been an appreciable mixture of these channels. The lower energy state that originally was formed from $\bar{B} N$ alone, now is still dominated by the $\bar{B} N$ channel but with an appreciable mixture of $\pi \Sigma_b$ and $\eta \Lambda_b$. On the other hand, the higher energy state is shown to be dominated by the $\pi \Sigma_b$ channel. However, the coupling to the $\bar{B} N$ state has been essential to obtain this state, since the single channel $\pi \Sigma_b$ does not produce it.

If one compares the energy of the lower energy state in Table \ref{tab:polepCoup0} with that of the single $\bar{B} N$ channel in Table \ref{tab:poleBbN0ng}, we can see that for $q_{max}=800\mev$ the effect of the coupled channels has been a reduction of about 40 MeV. Hence, even if suppressed, the coupled channels to the $\bar{B} N$ have a relevant role in the generation of states. In any case, we see that neither of the states found can qualify as the $\Lambda_b (5912)$, $\Lambda_b (5920)$. This is also the case for the higher energy state.

After this, we exploit another possibility, that these $\Lambda_b$ states come from $\bar{B}^* N$ and coupled channels. The $\bar{B}^* N$ can lead to two spins, $J^P = 1/2^-, \, 3/2^-$ and within the local hidden gauge approach the interaction is spin independent \citep{angelsvec}. Then we would get two degenerate states with spins $1/2$ and $3/2$. The $8\mev$ difference between $\Lambda_b (5912)$ and $\Lambda_b (5920)$ is small enough to fit into the category of degenerate. The degeneracy is broken with the mixture of the $V\,B$ and $P\,B$ states, which is done in Refs. \cite{javier,kanchan,kanchan2}, but for the heavier mesons this mixture is smaller \cite{xiaojuan}, which can explain the small difference between the masses of the two states. We shall come back to this point in the next two sections.

The binding of $\bar{B} N$ in Table \ref{tab:poleBbN0ng} for $q_{max} \sim 750 - 800 \mev$ is of the order of 300 MeV. While this is only $5\,\%$ of the total energy, it might surprise us that this amount is about three times bigger than the one obtained in Refs. \cite{wuzou,xiaooset} for hidden beauty baryons ($B \Sigma_b$ is the equivalent component), but this is easy to understand, both qualitatively and quantitatively. Indeed, in the exchange of light vectors between $\bar{B}$ and $N$, the nucleon has three light quarks, while in the exchange of a light vector between $B$ and $\Sigma_b$, the $\Sigma_b$ has only two light quarks. There are, hence, more chances to exchange a light vector between $\bar{B} N$ than in $B \Sigma_b$. More quantitatively, if we take $I=0$ for $\bar{B} N$ we have two components, $\bar{B}^0 (b \bar{d}) \, n(udd)$ and $\bar{B}^- (b \bar{u}) \, p(uud)$. We have two $d$ quarks from the $n$ to accommodate the exchange of a light $q \bar{q}$ in the first component and two $u$ quarks in the second component. If we take $B \Sigma_b$ in $I = 1/2$, which was found bound in Ref. \cite{xiaooset}, we have the components $B^0 (\bar{b}d) \, \Sigma_b^+ (uub)$ and $B^+ (\bar{b}u) \, \Sigma_b^0 (udb)$. In the first case we can not exchange a light $q \bar{q}$ vector and in the second case there is only one $u$ quark in the $\Sigma_b^0$ that can accommodate it. The strength of light vector exchange in $\bar{B} N, \ I=0$, should be much large than in $B \Sigma_b, \ I=1/2$. This is the case in practice since, comparing Table \ref{tab:vij0} of the present paper with Eqs. (2) and (12) of Ref. \cite{xiaooset}, we find that the relevant $C_{ij}$ coefficient is 3 for $\bar{B} N$ and 1 for $B \Sigma_b$. As a consequence, we have the about three times larger binding found here with respect to the one of Refs. \cite{wuzou,xiaooset}.

\section{Vector-baryon channels}

The transitions $VB \to VB$ for small momenta of the vector mesons have formally the same expressions as the corresponding $PB \to PB$ substituting the octet of pseudoscalars by the octet of vectors \cite{angelsvec}, with only one minor change to account for the $\phi$ and $\omega$ SU(3) structure, which is to replace each $\eta$ by $-\sqrt{2/3} ~\phi$ or $\sqrt{1/3} ~\omega$. The case of vector interaction with the decuplet of baryons is similar \cite{sarkarvec}. The Tables \ref{tab:vij0}, \ref{tab:vij1}, \ref{tab:vij1j3} are changed now to Tables \ref{tab:vij0j13}, \ref{tab:vij1j13}, \ref{tab:vij1j135}. Once again we penalize with a factor $1/4$ the transitions from a heavy vector to a light vector as we did before for the pseudoscalar mesons.
\begin{table}[H]
     \renewcommand{\arraystretch}{1.5}
     \setlength{\tabcolsep}{0.4cm}
\centering
\begin{tabular}{c|cccc}
$C_{ij}$ & $\bar{B}^* N$ &  $\rho \Sigma_b$ &  $\omega \Lambda_b$ & $\phi \Lambda_b$  \\
\hline
$\bar{B}^* N$ &  3 & $-\sqrt{\frac{3}{2}}$ & $\sqrt{\frac{3}{2}}$ & $-\sqrt{3}$   \\
$\rho \Sigma_b$ &   & 4  & 0  & 0 \\
$\omega \Lambda_b$ &  &  & 0 & 0  \\
$\phi \Lambda_b$&  &  & & 0
\end{tabular}
\caption{$C_{ij}$ coefficients for $\bar{B}^* N$ and coupled channels for $I=0$, and $J^P = 1/2^-, \, 3/2^-$.}
\label{tab:vij0j13}
\end{table}

\begin{table}[H]
     \renewcommand{\arraystretch}{1.5}
     \setlength{\tabcolsep}{0.4cm}
\centering
\begin{tabular}{c|ccccc}
$C_{ij}$ & $\bar{B}^* N$ &  $\rho \Sigma_b$ &  $\rho \Lambda_b$ & $\omega \Sigma_b$   & $\phi \Sigma_b$ \\
\hline
$\bar{B}^* N$ &  1 & $-1$ & $-\sqrt{\frac{3}{2}}$ & $-\sqrt{\frac{1}{2}}$ & 1   \\
$\rho \Sigma_b$ &   & 2  & 0  & 0  & 0  \\
$\rho \Lambda_b$ &  &  & 0 & 0  & 0   \\
$\omega \Sigma_b$ &  &  &  & 0  & 0    \\
$\phi \Sigma_b$ &  &  &  &  & 0
\end{tabular}
\caption{$C_{ij}$ coefficients for $\bar{B}^* N$ and coupled channels for $I=1$, and $J^P = 1/2^-, \, 3/2^-$.}
\label{tab:vij1j13}
\end{table}

\begin{table}[H]
     \renewcommand{\arraystretch}{1.5}
     \setlength{\tabcolsep}{0.4cm}
\centering
\begin{tabular}{c|cccc}
$C_{ij}$ & $\bar{B}^* \Delta$ &  $\rho \Sigma_b^*$ &  $\omega \Sigma_b^*$ & $\phi \Sigma_b^*$  \\
\hline
$\bar{B}^* \Delta$ &  4 & 1 & $\sqrt{2}$ & $-2$   \\
$\rho \Sigma_b^*$ &   & 2  & 0  & 0 \\
$\omega \Sigma_b^*$ &  &  & 0 & 0  \\
$\phi \Sigma_b^*$&  &  & & 0
\end{tabular}
\caption{$C_{ij}$ coefficients for $\bar{B}^* \Delta$ and coupled channels for $I=1$, and $J^P = 1/2^-, \, 3/2^-, \, 5/2^-$.}
\label{tab:vij1j135}
\end{table}

We take again the case of $I=0$ of Table \ref{tab:vij0j13} and show the results that obtain in Table \ref{tab:poleBsN0ng} for $\bar{B}^* N$ single channel, and in Table \ref{tab:polepCoup0j13} for coupled channels.
\begin{table}[H]
     \renewcommand{\arraystretch}{1.5}
     \setlength{\tabcolsep}{0.5cm}
\centering
\begin{tabular}{|c|c|c|c|c|}
\hline
$q_{max}$ & 700  &  750  &  800 & 850  \\
\hline
$V$  &  6033.1  &  5987.2  &  5939.0  &  5889.3   \\
\hline
\end{tabular}
\caption{Energies for a state of $\bar{B}^* N$ in $I=0$ as a function of $q_{max}$. (unit: MeV)}
\label{tab:poleBsN0ng}
\end{table}
\begin{table}[H]
     \renewcommand{\arraystretch}{1.5}
     \setlength{\tabcolsep}{0.3cm}
\centering
\begin{tabular}{|c|c|c|c|c|}
\hline
$q_{max}$ & 700  &  750  &  800 & 850  \\
\hline
  &  6019.2  &  5970.6  &  5919.8  &  5867.6   \\
\cline{2-5} \rb{$V$} &  $6364.6+i0.8$  & $6333.3+i0.8$ & $6303.0+i0.6$  & $6274.1+i0.3$  \\
\hline
\end{tabular}
\caption{Energies for a state in coupled channels $\bar{B}^* N$, $\rho \Sigma_b$, $\omega \Lambda_b$, $\phi \Lambda_b$ in $I=0$ as a function of $q_{max}$. (unit: MeV)}
\label{tab:polepCoup0j13}
\end{table}
Once again we see that the consideration of coupled channels leads to two states. In order to see the meaning of the states we calculate again the couplings to the different channels for $q_{max}=800\mev$, and the results are shown in Table \ref{tab:cou130}.
\begin{table}[ht]
     \renewcommand{\arraystretch}{1.2}
     \setlength{\tabcolsep}{0.3cm}
\centering
\begin{tabular}{ccccc}
\hline\hline
$5919.8+i0$ & $\bar{B}^* N$ & $\rho \Sigma_b$ & $\omega \Lambda_b$ & $\phi \Lambda_b$  \\
\hline
$g_i$ & $16.81$ & $1.04$ & $0.94$  & $1.33$ \\
$g_i\,G_i^{II}$ & $-22.01$ & $-5.46$ & $-6.16$ & $-5.67$  \\
\hline
$6303.0+i0.6$ & $\bar{B}^* N$ & $\rho \Sigma_b$ & $\omega \Lambda_b$ & $\phi \Lambda_b$  \\
\hline
$g_i$ & $0.37+i0.27$ & $5.14+i0.01$ & $0.15+i0.01$ & $0.21+i0.02$  \\
$g_i\,G_i^{II}$ & $-2.73-i0.27 \,$ & $\, -46.81-i0.13 \,$ & $\, -2.22-i0.22 \,$ & $\, -1.50-i0.15$  \\
\hline
\end{tabular}
\caption{The coupling constants to various channels for certain poles in the $J=1/2, \, 3/2,~I=0$ sector.} \label{tab:cou130}
\end{table}
There we can see that the state that couples strongly to $\bar{B}^* N$ is the one with lower energy. The higher energy state couples mostly to $\rho \Sigma_b$.

It is interesting to compare the results of Tables \ref{tab:poleBbN0ng} and \ref{tab:poleBsN0ng} for the states that couple mostly to $\bar{B} N$ and $\bar{B}^* N$. If we calculate with single channel we find a difference in energies between these two levels of 45 MeV, the same as between $m_{B^*}$ and $m_B$. However, when we include the coupled channels we see some changes. If we compare Tables \ref{tab:poleBsN0ng} and \ref{tab:polepCoup0j13} at $q_{max}=800\mev$, the effect of the coupled channels is a reduction of the mass of the lower state by about 20 MeV rather than 40 MeV in the case of $\bar{B} N$. The difference in the masses of the $\pi \Sigma_b$ or $\rho \Sigma_b$ is one of the reasons for it, but also the interaction of these two channels is different. Indeed, the $VVV$ vertices or $PPV$ vertices go as the sum of the external energies, as we saw, but now we have the much larger energy of the $\rho$ instead of the energy of the $\pi$.

\section{Breaking the $J=1/2^-, \, 3/2^-$ degeneracy in the $\bar{B}^* N$ sector}
\label{sec:mixing}

\begin{figure}[tb]
\epsfig{file=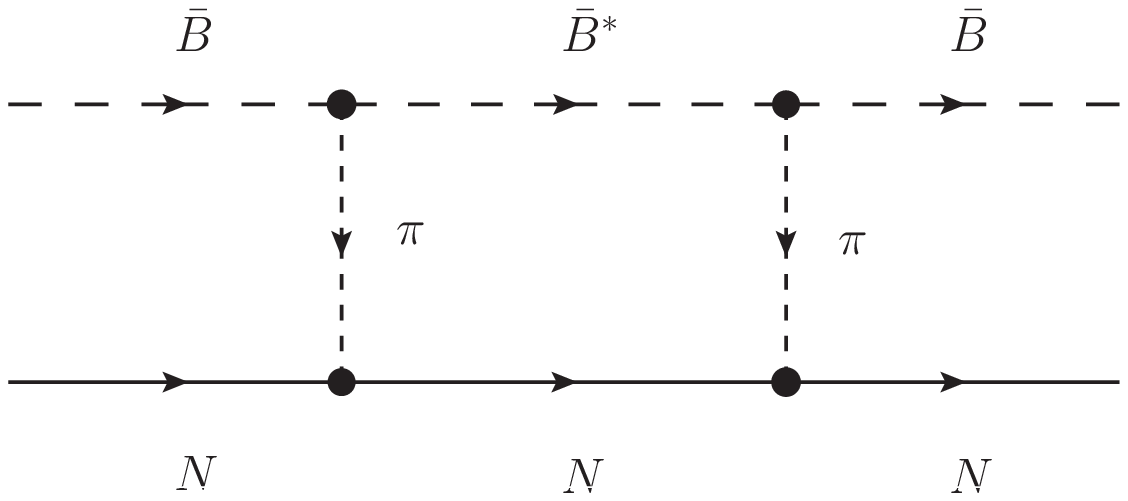, width=7cm} \epsfig{file=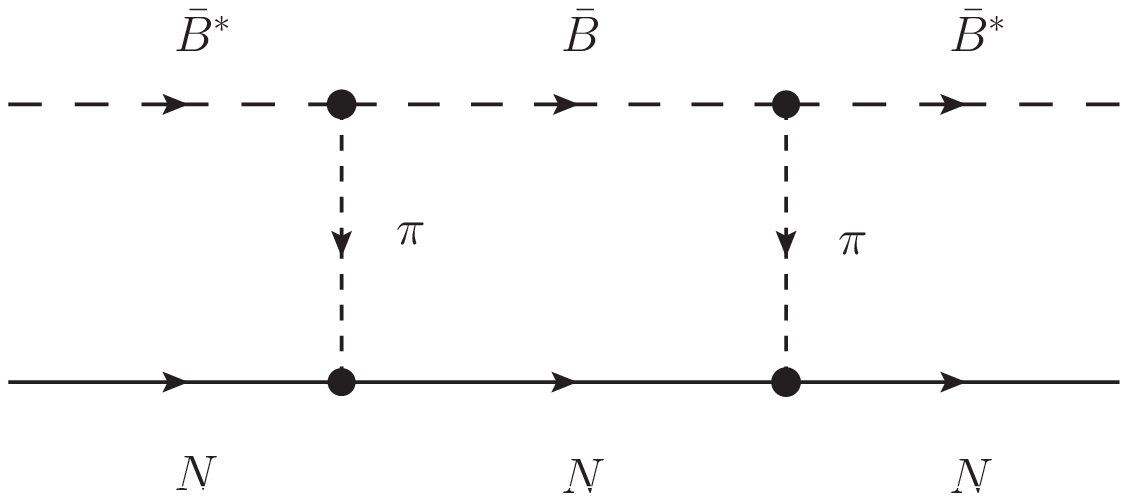, width=7cm}
\caption{Diagrammatic representation of the $\bar{B}^* N$ in intermediate state (left) and the $\bar{B} N$ in intermediate state (right).}\label{fig:bbbox}
\end{figure}

\begin{figure}[tb]
\epsfig{file=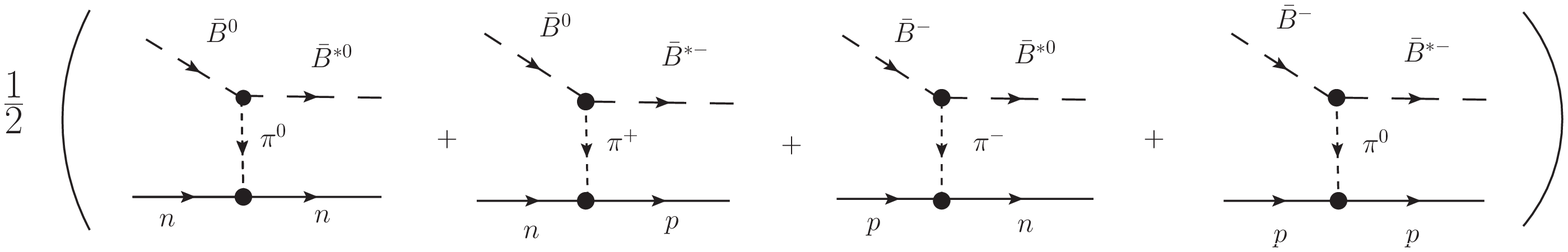, width=16cm}
\caption{Diagrammatic representation of the transition $\bar{B} N \to \bar{B}^* N$ in $I=0$.}\label{fig:bbboxhalf}
\end{figure}

In this section we shall break the degeneracy of the $1/2^-, \, 3/2^-$ states of the $\bar{B}^* N$ sector. For this purpose we follow the approach of Ref. \cite{javier} and mix states of $\bar{B}^* N$ and $\bar{B} N$ in both sectors. We test first that in the coupled channels like the $\bar{B}^* N$ sector, the important contribution comes from $\bar{B}^* N \to X \to \bar{B}^* N$, where $X$ stands for the other coupled channels. The extra interaction of the $X$ channels among themselves is negligible compared to that of the dominant $\bar{B}^* N$ channel, because of the big value of the $\bar{B}^*$ energy entering in the interaction. This means that it is sufficient to evaluate the contribution of the box diagrams of Fig. \ref{fig:bbbox},
in analogy to the box diagrams evaluated in Ref. \cite{javier}, and add this contribution, $\delta V$, to the $\bar{B} N$ or $\bar{B}^* N$ potential. Using the doublets of isospin $(B^+, \, B^0)$, $(\bar B^0, \, -B^-)$ the $\Lambda_c$ state in the $\bar{B} N$ basis is given by
\begin{equation}
|\bar{B} N, \, I=0 \rangle = \frac{1}{\sqrt{2}} (|\bar{B}^0 n \rangle + |B^- p \rangle),
\end{equation}
and analogously for $\bar{B}^* N$. The $\bar{B} N \to \bar{B}^* N$ transition in $I=0$ is given by the diagrams of Fig. \ref{fig:bbboxhalf}.

The $V \, P \, \pi$ vertex in SU(3) is given by the Lagrangian
\begin{equation}
{\cal L}_{VPP} = -ig ~\langle [P,\partial_{\mu}P]V^{\mu}\rangle, \label{eq:vpp}
\end{equation}
where $P,\ V^\mu$ are the ordinary meson octet and vector nonet SU(3) matrix of the corresponding fields
\begin{eqnarray}
P &=& \left(
\begin{array}{ccc}
\frac{\pi^0}{\sqrt{2}}+\frac{\eta_8}{\sqrt{6}}  & \pi^+     & K^{+}   \\
\pi^-      & -\frac{\pi^0}{\sqrt{2}}+\frac{\eta_8}{\sqrt{6}} & K^{0}   \\
K^{-}      & \bar{K}^{0}       & -\frac{2\eta_8}{\sqrt{6}}  \\
\end{array}
\right) \ ,  \\
V_\mu &=& \left(
\begin{array}{ccc}
\frac{\rho^0}{\sqrt{2}}+\frac{\omega}{\sqrt{2}} & \rho^+ & \quad K^{*+}  \\
 \rho^{-} & -\frac{\rho^0}{\sqrt{2}} + \frac{\omega}{\sqrt{2}} & K^{*0} \\
  K^{*-} & \bar K^{*0} & \phi \\
\end{array}
\right)_\mu \ .
\end{eqnarray}
and $g=m_V/2f_\pi$ with $m_V \approx 780 \mev$, $f_\pi=93$~MeV. One can extend the Lagrangian Eq. \eqref{eq:vpp} to the SU(4) space, as done in Ref. \cite{xiaoaltug}, but it is unnecessary. It is more intuitive and rigorous to follow the derivation below, which allows us to directly connect with the results of heavy quark spin-flavour symmetry \cite{Wise:1992hn}. Indeed, all we need to do is to invoke that the leading terms correspond to light meson exchange, in which case the heavy quark plays the role of a spectator at the quark level.

\begin{figure}[tb]
\epsfig{file=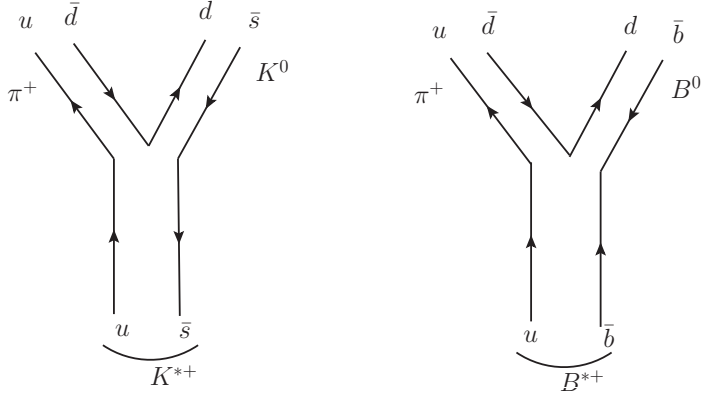, width=10cm}
\caption{Diagram of the transition $K^{*+} \to K^0 \pi^+$ (left) and $B^{*+} \to B^0 \pi^+$ (right).}\label{fig:tran}
\end{figure}

Let us then compare the $K^{*+} \to K^0 \pi^+$ and $B^{*+} \to B^0 \pi^+$ transitions as shown in Fig. \ref{fig:tran}.
As we can see in the figure, the transitions are identical and governed by the light quarks, with the $\bar{s}$ quark in $K^{*+}$ and $\bar{b}$ quark in $B^{*+}$ playing the role of a spectator. The transition amplitudes are thus identical at the quark microscopic level, but we must take into account that when used at the macroscopic level of the $K^{*+}$ or $B^{*+}$ there are normalization factors $(2 \omega)^{-1/2}$ which are different for the $K^{*+}$, $K^0$ or $B^{*+}$, $B^0$ fields. This is taken easily into account by constructing the $S$ matrix at the macroscopic level. At the microscopic level we have (we follow Mandl + Shaw normalization of the fields \cite{mandl})
\begin{equation}
S^{mic} = 1 - i t \sqrt{\frac{2m_L}{2E_L}} \sqrt{\frac{2m'_L}{2E'_L}} \sqrt{\frac{1}{2\omega_\pi}} \frac{1}{{\cal V}^{3/2}} \, (2\pi)^4 \delta(P_{in} - P_{out}), \label{eq:mic}
\end{equation}
with $m_L$, $E_L$, $m'_L$, $E'_L$ the masses (constituent) of the incoming and outgoing light quarks, ${\cal V}$ the volume of the box where states are normalized to unity, and $\omega_\pi$ the pion energy. At the macroscopic level we have for the $K^{*+}$ and $B^{*+}$
\begin{eqnarray}
S^{mac}_{K^*} &=& 1 - i t_{K^*} \frac{1}{\sqrt{2\omega_{K^*}}} \frac{1}{\sqrt{2\omega_K}} \frac{1}{\sqrt{2\omega_\pi}} \frac{1}{{\cal V}^{3/2}} \, (2\pi)^4 \delta(P_{in} - P_{out}), \label{eq:mick}  \\
S^{mac}_{B^*} &=& 1 - i t_{B^*} \frac{1}{\sqrt{2\omega_{B^*}}} \frac{1}{\sqrt{2\omega_B}} \frac{1}{\sqrt{2\omega_\pi}} \frac{1}{{\cal V}^{3/2}} \, (2\pi)^4 \delta(P_{in} - P_{out}). \label{eq:micb}
\end{eqnarray}
These considerations are common place in the study of three body systems in the fixed center approximation \cite{Roca:2010tf,Bayar:2011qj}. Eqs. \eqref{eq:mic}, \eqref{eq:mick}, \eqref{eq:micb} allow one to relate $t_{B^*}$ and $t_{K^*}$ with the macroscopic $t$ amplitude, but since we have $t_{K^*}$ given by the effective Lagrangian of Eq. \eqref{eq:vpp}, we can obtain $t_{B^*}$ in terms of $t_{K^*}$ by means
\begin{equation}
\frac{t_{B^*}}{t_{K^*}} \equiv \frac{\sqrt{m_{B^*}m_B}}{\sqrt{m_{K^*}m_K}} \simeq \frac{m_{B^*}}{m_{K^*}}. \label{eq:ratio}
\end{equation}
For a $B^*$ at rest, as we shall assume in our evaluations, $t$ is proportional to $\vec{\epsilon} \,\cdot\, \vec{q}$, with $\vec{q}$ the pion momentum and $\vec{\epsilon}$ the polarization vector of the vector meson (corrections of order $|\vec{p}_{B^*}|/m_{B^*}$ coming next can be safety neglected). It is interesting to compare what we get in our approach to the results of Ref. \cite{Wise:1992hn}. In Ref. \cite{Wise:1992hn} the width for $B^{*+} \to B^0 \pi^+$ (or $D^{*+} \to D^0 \pi^+$ ) is given by
\begin{equation}
\Gamma = \frac{g_H^2}{6 \pi \tilde{f}^2_\pi} |\vec{p}_\pi|^3,
\end{equation}
with $g_H$ the coupling appearing in the heavy hadron Lagrangian and $\tilde{f}_\pi = \sqrt{2} f_\pi$. For the same amplitude our approach, considering Eq. \eqref{eq:ratio}, is given by
\begin{equation}
\Gamma = \frac{1}{6 \pi} \frac{1}{m^2_{B^*}} g^2 \big( \frac{m_{B^*}}{m_{K^*}} \big)^2  |\vec{p}_\pi|^3.
\end{equation}
By taking $g^2/m^2_{K^*} = (m_V/2f_\pi m_{K^*})^2 \equiv \frac{1}{4 f^2_\pi}$, we have the relationship
\begin{equation}
\frac{g_H^2}{2} \equiv \frac{1}{4}; \quad g_H = \frac{1}{\sqrt{2}}.
\end{equation}
The same result would appear if we use another heavy vector decay like $D^*$. Our approach, with the consideration of the field normalizations leads to a $g_H$ independent of flavour and furthermore provides a value for it of $(\sqrt{2})^{-1}$. This value is in good agreement with the latest lattice QCD result \cite{Flynn:2013kwa} for the $B^* \to B \pi$ decay
\begin{equation}
g_H = 0.57 \pm 0.1.
\end{equation}
After this reformulation of the essence of the heavy quark symmetry, let us give a step forward and see what happens for the exchange of light vector mesons in the local hidden gauge approach. In Fig. \ref{fig:f1} the $PPV$ (or $VVV$) upper vertex gives rise to $\omega_p + \omega'_p$ (the factor $2 \sqrt{s} -M_{B1}- M_{B2}$ of Eq. \eqref{eq:vij} is the relativistic version of this magnitude). If we compare now $\bar K N \to \bar K N$ with $\bar B N \to \bar B N$, the ratio of amplitudes is given again by Eq. \eqref{eq:ratio}, substituting the energies by those involved here
\begin{equation}
\frac{t'_B}{t'_K} = \frac{\omega_B}{\omega_K},
\end{equation}
but now $t'_K$ is proportional to $\omega_K + \omega'_K$ in the local hidden gauge approach, and thus
\begin{equation}
t'_B \propto \frac{\omega_B}{\omega_K} \cdot 2 \omega_K \equiv 2 \omega_B.
\end{equation}
Hence, the $t'_B$ amplitude in this case, is the same one as $t'_K$ but substituting $\omega_K$ by $\omega_B$. This is exactly what the extension of the local hidden gauge approach to the $B$ sector gives if one simply substitutes the $s$ quark by the $b$ quark in the SU(3) Lagrangian \cite{xiaoaltug}, and it avoids having to invoke SU(4) arguments as often is done to justify this result.

\begin{figure}[tb]
\epsfig{file=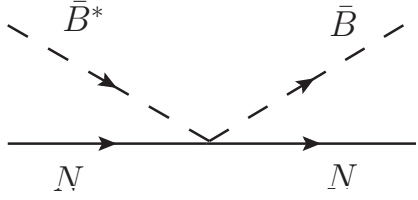, width=6cm}
\caption{Diagram of the Kroll Ruderman term.}\label{fig:kroll}
\end{figure}

\begin{figure}[tb]
\epsfig{file=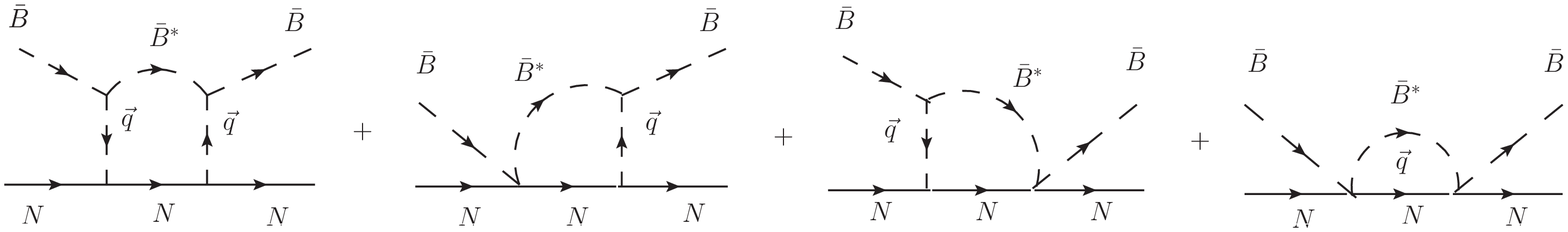, width=16cm}
\caption{All of the diagrams for the $\bar{B}^* N$ in the intermediate state.}\label{fig:bbboxtot}
\end{figure}

Now we come back to the evaluation of the box diagrams of Fig. \ref{fig:bbbox}. The vertex for the $I=0$ transition $\bar{B} N \to \bar{B}^* N$ of Fig. \ref{fig:bbboxhalf}, considering the Yukawa coupling for the $\pi NN$ vertex is given by
\begin{equation}
-it = - \frac{3}{\sqrt{2}} g \frac{m_{B^*}}{m_{K^*}} (q + P_{in})_\mu \epsilon^\mu \frac{1}{q^2 - m^2_\pi} \frac{D+F}{2f_\pi} \vec{\sigma}\,\cdot\, \vec{q}, \label{eq:tbox}
\end{equation}
with $D=0.75$ and $F=0.51$ \cite{Borasoy:1998pe}, and since $P_{in} = q + P_{out}$ and $P_{out} \cdot \epsilon =0$ plus $\epsilon^0 \approx 0$, we get effectively
\begin{equation}
-it = \frac{6}{\sqrt{2}} g \frac{m_{B^*}}{m_{K^*}} \vec{q}\,\cdot\, \vec{\epsilon} \frac{1}{q^2 - m^2_\pi} \frac{D+F}{2f_\pi} \vec{\sigma}\,\cdot\, \vec{q}.
\end{equation}
In addition to the pion exchange of Fig. \ref{fig:bbboxhalf}, we have the Kroll Ruderman contact term, depicted in Fig. \ref{fig:kroll}.
Following Refs. \cite{javier,Carrasco:1989vq}, in order to get the Kroll Ruderman term, we must substitute in Eq. \eqref{eq:tbox} $\epsilon_\mu (q + P_{in})^\mu \frac{1}{q^2 - m^2_\pi} \vec{\sigma}\,\cdot\, \vec{q}$ by $- \vec{\sigma}\,\cdot\, \vec{q}$. Then, we must evaluate the diagrams of Fig. \ref{fig:bbboxtot}
and we obtain
\begin{equation}
\delta V = \delta V^{PP} + 2 \delta V^{PC} + \delta V^{CC}, \label{eq:delv}
\end{equation}
where
\begin{eqnarray}
-i\delta V^{PP} &=& \int \frac{d^4 q}{(2\pi)^4} \Big( \frac{m_{B^*}}{m_{K^*}} \Big)^2 \, g \big( \frac{6}{\sqrt{2}} \vec{\epsilon} \,\cdot\,\vec{q} \frac{1}{q^{0\,2} - \vec{q}\,^2 - m^2_\pi} \frac{D+F}{2f_\pi} \vec{\sigma}\,\cdot\, \vec{q} \big) \nonumber \\
&& \times (-g) \big( \frac{6}{\sqrt{2}} \vec{\epsilon} \,\cdot\,\vec{q} \frac{1}{q^{0\,2} - \vec{q}\,^2 - m^2_\pi} \frac{D+F}{2f_\pi} \vec{\sigma}\,\cdot\, \vec{q} \big) \nonumber \\
&& \times i \frac{1}{2 \omega_{B^*}(\vec{q}\,)} \frac{1}{P^0_{in}- q^0 - \omega_{B^*}(\vec{q}\,) + i \epsilon} i \frac{M_N}{E_N(\vec{q}\,)} \frac{1}{K^0_{in}+q^0 - E_N(\vec{q}\,) + i \epsilon}, \label{eq:delvpp} \\
-i\delta V^{PC} &=& \int \frac{d^4 q}{(2\pi)^4} \Big( \frac{m_{B^*}}{m_{K^*}} \Big)^2 \, g \big( \frac{3}{\sqrt{2}}  \frac{D+F}{2f_\pi} \vec{\sigma}\,\cdot\, \vec{\epsilon}  \big) (-g) \big( \frac{6}{\sqrt{2}} \vec{\epsilon} \,\cdot\,\vec{q} \frac{1}{q^{0\,2} - \vec{q}\,^2 - m^2_\pi} \frac{D+F}{2f_\pi} \vec{\sigma}\,\cdot\, \vec{q} \big) \nonumber \\
&& \times i \frac{1}{2 \omega_{B^*}(\vec{q}\,)} \frac{1}{P^0_{in}- q^0 - \omega_{B^*}(\vec{q}\,) + i \epsilon} i \frac{M_N}{E_N(\vec{q}\,)} \frac{1}{K^0_{in}+q^0 - E_N(\vec{q}\,) + i \epsilon}, \label{eq:delvpc} \\
-i\delta V^{CC} &=& \int \frac{d^4 q}{(2\pi)^4} \Big( \frac{m_{B^*}}{m_{K^*}} \Big)^2 \, g \big( \frac{3}{\sqrt{2}}  \frac{D+F}{2f_\pi} \vec{\sigma}\,\cdot\, \vec{\epsilon}  \big) (-g) \big( \frac{3}{\sqrt{2}}  \frac{D+F}{2f_\pi} \vec{\sigma}\,\cdot\, \vec{\epsilon}  \big) \nonumber \\
&& \times i \frac{1}{2 \omega_{B^*}(\vec{q}\,)} \frac{1}{P^0_{in}- q^0 - \omega_{B^*}(\vec{q}\,) + i \epsilon} i \frac{M_N}{E_N(\vec{q}\,)} \frac{1}{K^0_{in}+q^0 - E_N(\vec{q}\,) + i \epsilon}.  \label{eq:delvcc}
\end{eqnarray}
Eqs. \eqref{eq:delv}, \eqref{eq:delvpp}, \eqref{eq:delvpc}, \eqref{eq:delvcc} can be further simplified considering that $(\vec{\sigma}\,\cdot\, \vec{q}\,)\, (\vec{\sigma}\,\cdot\, \vec{q}\,) = \vec{q}\,^2$ and resorting to symmetry properties
\begin{equation}
q_i q_j \to \frac{1}{3} \vec{q}\,^2 \delta_{ij}, \label{eq:qiqj}
\end{equation}
and further considering that, as shown in Ref. \cite{javier}
\begin{equation}
\langle PN |\vec{\sigma}\,\cdot\, \vec{\epsilon}\,| VN \rangle \equiv \sqrt{3} \delta_{J,1/2}.
\end{equation}
Thus in the $\bar{B} N \to \bar{B}^* N \to \bar{B} N$ transition, in $L=0$, we have $J=1/2$ and the Kroll Ruderman term contributes together with the pion in flight term (pion exchange).

In the $\bar{B}^* N \to \bar{B} N \to \bar{B}^* N$, we can have $L=0, \, 2$ in the intermediate $\bar{B} N$ states. Now in $\bar{B}^* N$ we can have $J=1/2, \, 3/2$. If $J=1/2$ then $L=0$ for the $\bar{B} N$ intermediate states and both the pion in flight and Kroll Ruderman terms contribute. The formulas are then formally the same as for the $\bar{B} N \to \bar{B}^* N \to \bar{B} N$ box diagram, changing appropriately the masses. However, if we have $J=3/2$, then the Kroll Ruderman term does not contribute, and the pion in flight, $\delta V^{PP}$, with $L=2$ contributes the same as for $L=0$. Furthermore, before we had the vertices $(\vec{\epsilon} \,\cdot\, \vec{q} \,) \, (\vec{\epsilon} \,\cdot\, \vec{q}\,)$ and we must sum over polarizations of the $\vec{\epsilon}$, so one get the factor $\vec{q}\,^2$, but in the $\bar{B}^* N \to \bar{B} N \to \bar{B}^* N$ transition we have $(\vec{\epsilon} \,\cdot\, \vec{q} \,)\, (\vec{\epsilon}\,' \,\cdot\, \vec{q}\,'\,)$ and $\vec{\epsilon}$, $\vec{\epsilon}\,'$ are the external polarization vectors. The symmetry of the integral allows us to use Eq. \eqref{eq:qiqj} and then we get $\frac{1}{3} \vec{q}\,^2 \, \vec{\epsilon} \,\cdot\, \vec{\epsilon}\,'$, so we get a factor $\frac{1}{3}$ in the $PP$ term of Eq. \eqref{eq:delvpp}. The scalar $\vec{\epsilon} \,\cdot\, \vec{\epsilon}\,'$ factor tells us that this term is the same for $J=1/2, \, 3/2$.

One further step to simplify the equations comes from performing the $q^0$ integration analytically. After all this is done we obtain the following results for the $\bar{B}^* N \to \bar{B} N \to \bar{B}^* N$ box diagram
\begin{eqnarray}
J &=& 1/2: \quad \delta V = FAC \big( \frac{\partial}{\partial m^2_\pi} I_1' + 2\, I_2' + I_3' \big), \\
J &=& 3/2: \quad \delta V = FAC \big( \frac{\partial}{\partial m^2_\pi} I_1' \big),
\end{eqnarray}
where
\begin{eqnarray}
I_1' &=& \int \frac{d^3 q}{(2\pi)^3} \frac{4}{3} \vec{q}\,^4 \frac{1}{2 \omega_B (\vec{q}\,)} \frac{M_N}{E_N (\vec{q}\,)} \frac{Num}{Den}, \label{eq:i1} \\
I_2' &=& \int \frac{d^3 q}{(2\pi)^3} 2 \vec{q}\,^2 \frac{1}{2 \omega_B (\vec{q}\,)} \frac{M_N}{E_N (\vec{q}\,)} \frac{Num}{Den}, \label{eq:i2}  \\
I_3' &=& \int \frac{d^3 q}{(2\pi)^3} \frac{3}{2 \omega_B (\vec{q}\,)} \frac{M_N}{E_N (\vec{q}\,)} \frac{1}{P^0_{in} + K^0_{in} -E_N (\vec{q}\,) - \omega_B (\vec{q}\,) + i \epsilon},  \label{eq:i3}
\end{eqnarray}
with
\begin{eqnarray}
FAC &=& \frac{9}{2} g^2 \Big( \frac{m_{B^*}}{m_{K^*}} \Big)^2 \Big( \frac{F+D}{2f_\pi} \Big)^2, \\
Num &=& K^0_{in} - E_N (\vec{q}\,) - 2 \omega_\pi (\vec{q}\,) - \omega_B (\vec{q}\,) + P^0_{in}, \label{eq:num} \\
Den &=& 2 \omega_\pi (\vec{q}\,) [P^0_{in} - \omega_\pi (\vec{q}\,) - \omega_B (\vec{q}\,) + i \epsilon] [K^0_{in} - E_N (\vec{q}\,) - \omega_\pi (\vec{q}\,) + i \epsilon] \nonumber \\
 && \times [P^0_{in} + K^0_{in} - E_N (\vec{q}\,) - \omega_B (\vec{q}\,) + i \epsilon], \label{eq:den}
\end{eqnarray}
with $P^0_{in}$, $K^0_{in}$ the incoming $\bar{B}^*, \, N$ energies.

For the case of the $\bar{B} N \to \bar{B}^* N \to \bar{B} N$ box diagram we only have the $J=1/2$ case, and the formula is like the former one for $J=1/2$ exchanging accordingly the masses of $B^* \leftrightarrow B$, and considering the factor 3 extra in $V^{PP}$. Thus, we have
\begin{equation}
J = 1/2: \quad \delta V = FAC \big( \frac{\partial}{\partial m^2_\pi} I_1 + 2\, I_2 + I_3 \big)
\end{equation}
where
\begin{eqnarray}
I_1 &=& \int \frac{d^3 q}{(2\pi)^3} 4 \vec{q}\,^4 \frac{1}{2 \omega_{B^*} (\vec{q}\,)} \frac{M_N}{E_N (\vec{q}\,)} \frac{Num}{Den}, \label{eq:i1a}  \\
I_2 &=& \int \frac{d^3 q}{(2\pi)^3} 2 \vec{q}\,^2 \frac{1}{2 \omega_{B^*} (\vec{q}\,)} \frac{M_N}{E_N (\vec{q}\,)} \frac{Num}{Den}, \label{eq:i2a}   \\
I_3 &=& \int \frac{d^3 q}{(2\pi)^3} \frac{3}{2 \omega_{B^*} (\vec{q}\,)} \frac{M_N}{E_N (\vec{q}\,)} \frac{1}{P^0_{in} + K^0_{in} -E_N (\vec{q}\,) - \omega_{B^*} (\vec{q}\,) + i \epsilon},   \label{eq:i3a}
\end{eqnarray}
with $Num,\ Den$ having the same expressions but in terms of the proper energies and masses. In the Yukawa vertex it is customary to include a monopole form factor to agree with the $NN$ peripheral partial waves \cite{Machleidt:1987hj} and thus we introduce a factor
\begin{equation}
FF(\vec{q}\,) = \Big( \frac{\Lambda^2}{\Lambda^2 + \vec{q}\,^2} \Big)^2,
\end{equation}
with $\Lambda \simeq 1 \gev$, which we include in the integrands of Eqs. \eqref{eq:i1}, \eqref{eq:i2}, \eqref{eq:i3},  \eqref{eq:i1a}, \eqref{eq:i2a}, \eqref{eq:i3a}.

To go to the second Riemann sheet with the box contribution $\delta V$, we can do a similar thing as in Eq. \eqref{eq:rs2}. Yet, when $\sqrt{s}$ is real $G_l^{II} \equiv (G_l^I)^*$ and this is also the case, quite accurately, when we are close to the real axis. In view of this, and the small contribution of the intermediate $\bar{B} N$ states to the width, we use the prescription, $\delta V \to (\delta V)^*$ to go to the second Riemann sheet. In practical, this is equivalent to changing $+i \epsilon \to -i \epsilon$ in the factors of $Den$ of Eq. \eqref{eq:den} \footnote{In this case only the last factor of Eq. \eqref{eq:den} can produce a singularity and the analytical structure is like the one $G$. For the $\bar{B}^* \Delta \to \bar{B} N \to \bar{B}^* \Delta$ case, that we shall study later, also the second last factor can lead to a singularity from $\Delta \to \pi N$, but the binding of the $\bar{B}^* \Delta$ state does not give much phase space for it. Hence, we apply the same rule. At the end we check that the width that comes out from the complex pole coincides with that of $|T|^2$ in the real axis.}.

Once the formalism has been described, we show the results of including $\delta V$ in the approach in Tables \ref{tab:polepCoup0j13box}, \ref{tab:polepCoup0box}.
\begin{table}[H]
     \renewcommand{\arraystretch}{1.5}
     \setlength{\tabcolsep}{0.2cm}
\centering
\begin{tabular}{|c|c|c|c|c|c|}
\hline
$q_{max}$ & 700  & 750  & 776 &  800 & 850  \\
\hline
  &  $5991.9+i0$  & $5939.2+i0$  & $5910.7+i0$  &  $5884.0+i0$  &  $5827.4+i0$   \\
\cline{2-6} \rb{$J=1/2$} &  $6364.2+i1.4$  & $6332.6+i1.4$ & $6316.6+i1.4$ & $6301.1+i1.4$  & $6273.0+i1.2$  \\
\hline
  &  $5998.8+i0$  & $5948.0+i0$  & $5920.7+i0$  &  $5895.1+i0$  &  $5840.9+i0$   \\
\cline{2-6} \rb{$J=3/2$} &  $6363.5+i2.1$  & $6331.7+i2.0$ & $6315.7+i1.9$ & $6301.2+i1.7$  & $6272.1+i1.2$  \\
\hline
\end{tabular}
\caption{Poles with box diagram in coupled channels $\bar{B}^* N$, $\rho \Sigma_b$, $\omega \Lambda_b$, $\phi \Lambda_b$ in $I=0$ as a function of $V$ and $q_{max}$. (unit: MeV)}
\label{tab:polepCoup0j13box}
\end{table}

\begin{table}[H]
     \renewcommand{\arraystretch}{1.5}
     \setlength{\tabcolsep}{0.2cm}
\centering
\begin{tabular}{|c|c|c|c|c|c|}
\hline
$q_{max}$ & 700  &  750  &  776  &  800 & 850  \\
\hline
  &  $5902.6+i0$  & $5850.1+i0$ & $5820.9+i0$ & $5793.3+i0$  & $5734.4+i0$  \\
\cline{2-6} \rb{$J=1/2$} &  $5985.6+i29.1$  & $5974.1+i26.8$ & $5969.5+i24.6$ & $5965.8+i22.3$  & $5959.5+i17.3$ \\
\hline
\end{tabular}
\caption{Poles with box diagram in coupled channels $\bar{B} N$, $\pi \Sigma_b$, $\eta \Lambda_b$ in $I=0$ as a function of $V$ and $q_{max}$. (unit: MeV)}
\label{tab:polepCoup0box}
\end{table}

It is interesting to compare the results of Table \ref{tab:polepCoup0j13box} with those of Table \ref{tab:polepCoup0j13}. At 750 MeV the box diagram reduces the mass of the state by about 30 MeV for $J=1/2$, and 20 MeV for $J=3/2$. We can see that the value of the masses is  rather sensitive to the value $q_{max}$ used. However, it is interesting to remark that the splitting of energies between the $J=1/2$ and $J=3/2$ levels is about 10 MeV, rather independent of the cutoff used. We can thus see that the mixing of $\bar{B}^* N$ and $\bar{B} N$ states leads naturally to two states, nearly degenerate in spin, only separated by about 10 MeV, like the $\Lambda_b(5912)$ and $\Lambda_b(5920)$. If we fine tune $q_{max}$ to get the right binding, we find $q_{max}=776\mev$, where the energy of the $J=1/2$ state is 5910 MeV, and the one of the $J=3/2$ state 5920 MeV. We also observe that the higher energy state, around 6300 MeV, has been practically not affected by the box diagram, which is most logical since this state couples weakly to $\bar{B}^* N$.

For the $\bar{B} N$ state of Table \ref{tab:polepCoup0box}, comparing it with the results of Table \ref{tab:polepCoup0}, the box diagram has reduced the energy by about 50 MeV at $q_{max}=750\mev$. The upper level energy is increased by about 15 MeV for this value of $q_{max}$.

\section{$I=1$ states}

With the cut off obtained to reproduce the mass of the $\Lambda_b(5912)$ state we proceed now to evaluate the states corresponding to Tables \ref{tab:vij1}, \ref{tab:vij1j3}, \ref{tab:vij1j13}, \ref{tab:vij1j135}, which are mostly bound states of $\bar{B} N \, (I=1, \, J^P = 1/2^-)$, $\bar{B} \Delta \, (I=1, \, J^P = 3/2^-)$, $\bar{B}^* N \, (I=1, \, J^P = 1/2^-, \, 3/2^-)$, $\bar{B}^* \Delta \, (I=1, \, J^P = 1/2^-, \, 3/2^-, \, 5/2^-)$. The results can be seen in Tables \ref{tab:polecou1}, \ref{tab:polecou2}, \ref{tab:polecou3}, \ref{tab:polecou4} with $q_{max}=776\mev$, together with the couplings to the different coupled channels.
\begin{table}[ht]
     \renewcommand{\arraystretch}{1.2}
\centering
\begin{tabular}{ccccc}
\hline\hline
$6002.8+i66.2$ & $\bar{B} N$ & $\pi \Sigma_b$ & $\pi \Lambda_b$ & $\eta \Sigma_b$  \\
\hline
$g_i$ & $5.69+i1.62$ & $1.80+i1.00$ & $0.39+i0.21$  & $0.37+i0.20$ \\
$g_i\,G_i^{II}$ & $-8.60-i4.12$ & $-72.70-i12.65$ & $-16.12+i5.73$ & $-2.95-i2.01$  \\
\hline
$6179.4+i61.4$ & $\bar{B} N$ & $\pi \Sigma_b$ & $\pi \Lambda_b$ & $\eta \Sigma_b$ \\
\hline
$g_i$ & $6.76-i3.03$ & $0.30+i0.80$ & $1.11-i0.03$ & $1.07-i0.03$  \\
$g_i\,G_i^{II}$ & $-21.71+i1.38 \,$ & $\, -30.00-i13.31 \,$ & $\, -14.96+i47.71 \,$ & $\, -12.74-i1.63$  \\
\hline
\end{tabular}
\caption{The coupling constants to various channels for certain poles in the $I=1$ sector of $\bar{B} N$ and coupled channels.} \label{tab:polecou1}
\end{table}

\begin{table}[ht]
     \renewcommand{\arraystretch}{1.2}
     \setlength{\tabcolsep}{0.3cm}
\centering
\begin{tabular}{cccc}
\hline\hline
$5971.9+i0$ & $\bar{B} \Delta$ & $\pi \Sigma_b^*$  & $\eta \Sigma_b^*$  \\
\hline
$g_i$ & $10.79$ & $0.61$ & $0.85$   \\
$g_i\,G_i^{II}$ & $-10.89$ & $-16.51$ & $-6.69$   \\
\hline
$6073.0+i77.2$  & $\bar{B} \Delta$ & $\pi \Sigma_b^*$  & $\eta \Sigma_b^*$ \\
\hline
$g_i$ & $7.67-i5.14$ & $1.43+i1.36$ & $0.81-i0.36$   \\
$g_i\,G_i^{II}$ & $-9.59+i4.83 \,$ & $\, -70.37-i20.89 \,$  & $\, -7.87+i2.27$  \\
\hline
\end{tabular}
\caption{The coupling constants to various channels for certain poles in the $I=1$ sector of $\bar{B} \Delta$ and coupled channels.} \label{tab:polecou2}
\end{table}

\begin{table}[ht]
     \renewcommand{\arraystretch}{1.2}
\centering
\begin{tabular}{cccccc}
\hline\hline
$6202.1+i0$ & $\bar{B}^* N$ & $\rho \Sigma_b$ & $\rho \Lambda_b$ & $\omega \Sigma_b$  & $\phi \Sigma_b$ \\
\hline
$g_i$ & $7.29$ & $1.22$ & $1.05$  & $0.59$   & $0.83$ \\
$g_i\,G_i^{II}$ & $-20.66$ & $-8.67$ & $-11.20$ & $-4.11$   & $-3.62$ \\
\hline
$6477.2+i5.0$ & $\bar{B}^* N$ & $\rho \Sigma_b$ & $\rho \Lambda_b$ & $\omega \Sigma_b$  & $\phi \Sigma_b$ \\
\hline
$g_i$ & $0.21-i0.62$ & $3.68-i0.08$ & $0.25+i0.12$ & $0.14+i0.07$ & $0.20+i0.10$   \\
$g_i\,G_i^{II}$ & $4.23+i2.05 \,$ & $\, -50.58+i0.19 \,$ & $\, -9.39+i2.89 \,$  & $\, -1.85-i0.94 \,$ & $\, -1.29-i0.64$  \\
\hline
\end{tabular}
\caption{The coupling constants to various channels for certain poles in the $I=1$ sector of $\bar{B}^* N$ and coupled channels.} \label{tab:polecou3}
\end{table}

\begin{table}[ht]
     \renewcommand{\arraystretch}{1.2}
      \setlength{\tabcolsep}{0.3cm}
\centering
\begin{tabular}{ccccc}
\hline\hline
$6049.2+i0$ & $\bar{B}^* \Delta$ & $\rho \Sigma_b^*$  & $\omega \Sigma_b^*$  & $\phi \Sigma_b^*$ \\
\hline
$g_i$ & $21.14$ & $0.85$ & $1.03$  & $1.46$   \\
$g_i\,G_i^{II}$ & $-22.11$ & $-4.70$ & $-5.66$ & $-5.30$  \\
\hline
$6491.9+i0$ & $\bar{B}^* \Delta$ & $\rho \Sigma_b^*$  & $\omega \Sigma_b^*$  & $\phi \Sigma_b^*$ \\
\hline
$g_i$ & $0.66$ & $3.76$ & $0.13$ & $0.18$   \\
$g_i\,G_i^{II}$ & $-2.19 \,$ & $\, -50.69 \,$ & $\, -1.71 \,$   & $\, -1.20$  \\
\hline
\end{tabular}
\caption{The coupling constants to various channels for certain poles in the $I=1$ sector of $\bar{B}^* \Delta$ and coupled channels.} \label{tab:polecou4}
\end{table}

With respect to their thresholds, the binding energies for the $\bar{B} N$ channel are now smaller than for $I=0$, as we anticipated in view of the smaller $C_{ij}$ coefficients, but for $\bar{B} \Delta$ the binding is bigger than for the $\bar{B} N$ state, as discussed earlier. We again see that we get two states in each one of the cases, but also notice that  there is more mixture of the states than for $I=0$. In the case of the $\bar{B} N$ channels, the lower state is clearly dominated by $\pi \Sigma_b$. For the $\bar{B} \Delta$ channels, the upper state is dominated by $\pi \Sigma_b^*$. For the $\bar{B}^* N$ channels, the lower state is dominated by $\bar{B}^* N$ and the higher one by $\rho \Sigma_b$. For the $\bar{B}^* \Delta$ channels, the lower state is dominated by $\bar{B}^* \Delta$ and the upper one by $\rho \Sigma_b^*$.

\section{Box diagram for $I=1$ states}

\begin{figure}[tb]
\epsfig{file=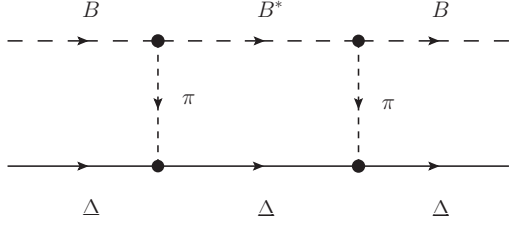, width=7cm}
\caption{Diagrammatic representation of the transition of $\bar{B} \Delta \to \bar{B}^* \Delta \to \bar{B} \Delta$.}\label{fig:bbdeldbox}
\end{figure}

We now evaluate the contribution of the box diagram to the $I=1$ states made from $\bar{B} N$, $\bar{B}^* N$,  $\bar{B} \Delta$, $\bar{B}^* \Delta$.\\

a) $\bar{B} N$, $I=1$:

The isospin $I=1$ state is now
\begin{equation}
| \bar{B} N; I=1, I_3=0 \rangle = \frac{1}{\sqrt{2}} \big( |\bar{B}^0 n \rangle - | \bar{B}^- p \rangle \big).
\end{equation}
The counting of isospin done before can be repeated and we simply find that a factor $\frac{3}{\sqrt{2}}$ gets converted in $\frac{1}{\sqrt{2}}$ in the $\bar{B} N \to \bar{B}^* N$ transition. We thus get a factor 9 smaller contribution than for $I=0$ from the box and we neglect it.\\

b) $\bar{B}^* N$, $I=1$:

We have the same suppression factor as before and we also neglect it.\\

c) $\bar{B} \Delta$, $I=1$:

The state of $\bar{B} \Delta$ with $I=1$ is given by
\begin{equation}
| \bar{B} \Delta; I=1, I_3=1 \rangle = \sqrt{\frac{3}{4}} |\bar{B}^- \Delta^{++} \rangle + \sqrt{\frac{1}{4}}| \bar{B}^0 \Delta^+ \rangle. \label{eq:bbardel}
\end{equation}
The diagram under contribution is now in Fig. \ref{fig:bbdeldbox}.
We must also substitute $\frac{f}{m_\pi} \vec{\sigma} \,\cdot\, \vec{q} \, \tau^\lambda$ in the case of nucleons by $\frac{f_\Delta}{m_\pi} \vec{S}_\Delta \cdot\, \vec{q} \, T^\lambda_\Delta$, where $\vec{S}_\Delta$, $\vec{T}_\Delta$ are the ordinary spin and isospin matrices of the $\Delta$.

We have \cite{manolo}
\begin{equation}
\frac{f_\Delta}{f}=\frac{4}{5}, \quad (\text{where} \quad \frac{f}{m_\pi}=\frac{F+D}{2f}).
\end{equation}
Using the appropriate Clebsch-Gordan coefficient for $\vec{T}_\Delta$, we find that the term corresponding the box in diagram Fig. \ref{fig:bbdeldbox}. is now given by
\begin{equation}
\delta V = FAC \frac{\partial \tilde{I}_1}{\partial m^2_\pi},
\end{equation}
where
\begin{equation}
\tilde{I}_1 = \frac{5}{9} \, \int \frac{d^3 q}{(2\pi)^3} 4 \vec{q}\,^4 \frac{1}{2 \omega_{B^*} (\vec{q}\,)} \frac{M_\Delta}{E_\Delta (\vec{q}\,)} \frac{Num}{Den},
\end{equation}
with $Num, \, Den$ the expressions of Eqs. \eqref{eq:num}, \eqref{eq:den} but putting the appropriate masses.\\

d) $\bar{B}^* \Delta$, $I=1$:

In this case we proceed as before, and everything is formulated in the same way but now $I_1' \to \tilde{I}_1'$, with
\begin{equation}
\tilde{I}_1' = \frac{5}{9} \, \int \frac{d^3 q}{(2\pi)^3} \frac{4}{3} \vec{q}\,^4 \frac{1}{2 \omega_B (\vec{q}\,)} \frac{M_\Delta}{E_\Delta (\vec{q}\,)} \frac{Num}{Den}.
\end{equation}
To reach this formula we have made an average over the spins of the initial $\Delta$, taking the same initial and final third spin component of the $\Delta$. This is in consonance with the fact that  since we have a reduction factor of about $1/2$, the splitting of spin levels is now smaller than for $\bar{B}^* N$ and accepting uncertainties larger than 5 MeV we do not worry about it. Consequently we do not evaluate the $I_2$, $I_3$, $I_2'$, $I_3'$ terms that produced the spin splitting.

\section{Further decay channels of $\bar{B} \Delta$ and $\bar{B}^* \Delta$}

\begin{figure}[tb]
\epsfig{file=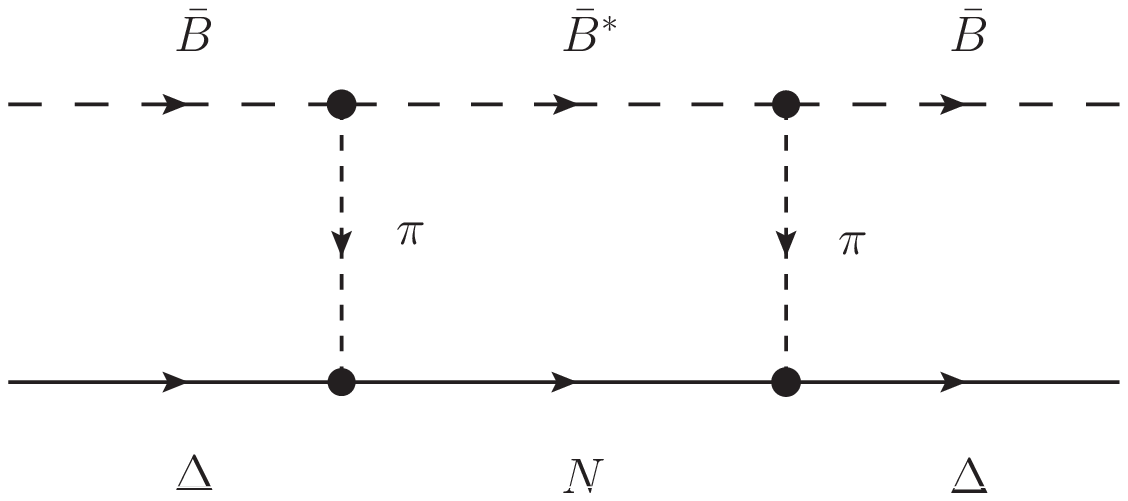, width=7cm} \epsfig{file=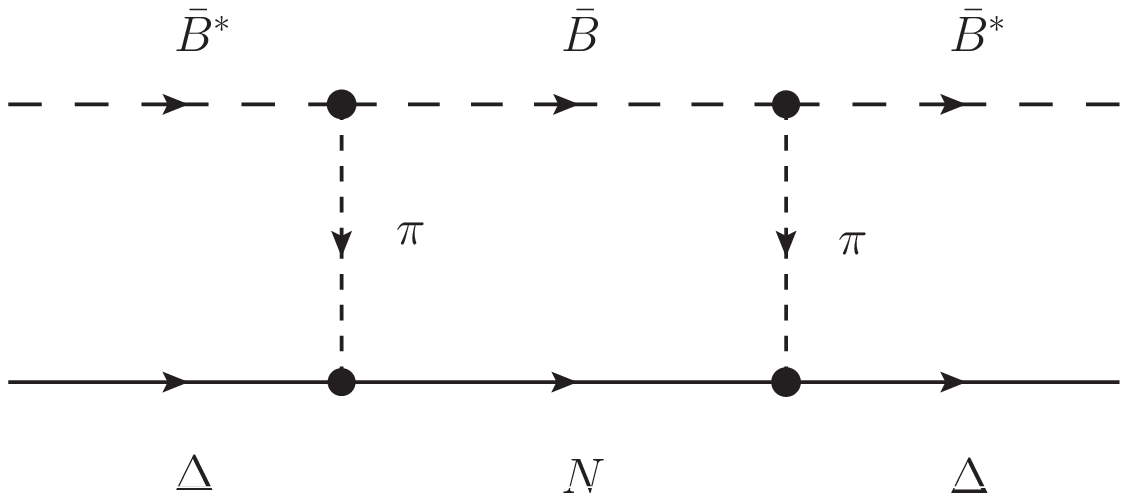, width=7cm}
\caption{Diagrammatic representation of the $\bar{B}^* N$ in intermediate state (left) and the $\bar{B} N$ in intermediate state (right).}\label{fig:bbdelbox}
\end{figure}

In this section we evaluate the box diagram corresponding to Figs. \ref{fig:bbdelbox}.
We thus consider the intermediate $\bar{B} N$ or $\bar{B}^* N$ channels. Should the binding $\bar{B} \Delta$ and $\bar{B}^* \Delta$ states be not bigger than the $\Delta$ and $N$ mass differences, there would provide decay channels of the states. In principle we should also consider the $\bar{B} \Delta$ and $\bar{B}^* \Delta$ intermediate states for the $\bar{B} N$ and $\bar{B}^* N$ states, but, considering the binding, these intermediate states are bout 600 MeV away in energy and we do not consider them.  The changes are also simple: we must substitute
\begin{equation}
\frac{f}{m_\pi} \vec{\sigma} \,\cdot\, \vec{q} \, \tau^\lambda \to \frac{f_{\pi N\Delta}}{m_\pi} \vec{S} \,\cdot\, \vec{q} \, T^\lambda,
\end{equation}
where now $\vec{S} \, (\vec{T})$ is transition spin (isospin) operator from spin (isospin) 3/2 to 1/2, with the normalization for $S_\mu^+$ in spherical basis
\begin{equation}
\langle 3/2 \, M |S_\mu^+| 1/2 \, m \rangle = {\cal C} (1/2 \; 1 \; 3/2; \; m_\mu \, M),
\end{equation}
with ${\cal C} (\cdot)$ a Clebsch-Gordan coefficient, and we have the property \cite{Oset:1981ih}
\begin{equation}
\sum_M S_i |M \rangle \langle M| S_j = \frac{2}{3} \delta_{ij} - i \frac{1}{3} \epsilon_{ijk} \sigma_{k},
\end{equation}
for $S_i$, $S_j$ in cartesian basis. Also from Ref. \cite{Oset:1981ih} we take $f_{\pi N\Delta}/f = 2.25$.  For the isospin transition, in addition to the $I=1$ $\bar{B} \Delta$ state of Eq. \eqref{eq:bbardel} we need
\begin{equation}
|\bar{B} N; I=1, I_3=1 \rangle = |\bar{B}^0 p \rangle.
\end{equation}
The $\pi^0 p \Delta^+$ vertex for $T_\Delta^\lambda$ gives us $\sqrt{2/3}$ and the $\pi^+ p \Delta^{++}$ gives us $(-1)$ (recall phase used $|\pi^+ \rangle = -|1,1 \rangle$).

Once again, making the average over $\Delta$ spins as before, we obtain the results for $\delta V_2$ given by the same formalism as before, substituting
\begin{equation}
I_1 \to \tilde{I}_1 = \frac{8}{81} \big( \frac{f_{\pi N \Delta}}{f} \big)^2 \int \frac{d^3 q}{(2\pi)^3} 4 \vec{q}\,^4 \frac{1}{2 \omega_{B^*} (\vec{q}\,)} \frac{M_N}{E_N (\vec{q}\,)} \frac{Num}{Den},
\end{equation}
for the $\bar{B} \Delta \to \bar{B}^* N \to \bar{B} \Delta$ process, and
\begin{equation}
I_1' \to \tilde{I}_1' = \frac{8}{81} \big( \frac{f_{\pi N \Delta}}{f} \big)^2 \int \frac{d^3 q}{(2\pi)^3} \frac{4}{3} \vec{q}\,^4 \frac{1}{2 \omega_B (\vec{q}\,)} \frac{M_N}{E_N (\vec{q}\,)} \frac{Num}{Den},
\end{equation}
for the $\bar{B}^* \Delta \to \bar{B} N \to \bar{B}^* \Delta$ process, with $Num$ and $Den$ given by Eqs. \eqref{eq:num}, \eqref{eq:den}, but substituting the masses by the appropriate ones.

We show our results with the contribution of box diagrams, seen in Figs. \ref{fig:bbdeldbox} and \ref{fig:bbdelbox}, in Tables \ref{tab:polepCoupbox1}, \ref{tab:polepCoupbox2} for the $\bar{B} \Delta$, $\bar{B}^* \Delta$ and their coupled channels.
\begin{table}[H]
     \renewcommand{\arraystretch}{1.5}
     \setlength{\tabcolsep}{0.3cm}
\centering
\begin{tabular}{|c|c|c|c|}
\hline
no box & $V+\delta V_1$   &  $V+\delta V_2$  &   $V+\delta V_1+\delta V_2$   \\
\hline
$5971.9+i0$  &  $5957.8+i0$  & $5949.4+i0$ & $5932.9+i0$   \\
\hline
$6073.0+i77.2$  &  $6069.1+i80.7$  & $6066.3+i81.7$ & $6063.8+i83.5$   \\
\hline
\end{tabular}
\caption{Poles with box diagram in $I=1$ sector of $\bar{B} \Delta$ and its coupled channels with $q_{max}=776\mev$: $\delta V_1$ is the $\bar{B}^* \Delta$ box, $\delta V_2$ is the $\bar{B}^* N$ box. (unit: MeV)}
\label{tab:polepCoupbox1}
\end{table}

\begin{table}[H]
     \renewcommand{\arraystretch}{1.5}
     \setlength{\tabcolsep}{0.3cm}
\centering
\begin{tabular}{|c|c|c|c|}
\hline
no box & $V+\delta V_1$   &  $V+\delta V_2$  &   $V+\delta V_1+\delta V_2$   \\
\hline
$6049.2+i0$  &  $6039.1+i0$  & $6032.2+i0$ & $6022.9+i0$   \\
\hline
$6491.9+i0$  &  $6491.4+i0$  & $6493.0+i1.0$ & $6491.7+i0.8$   \\
\hline
\end{tabular}
\caption{Poles with box diagram in $I=1$ sector of $\bar{B}^* \Delta$ and its coupled channels with $q_{max}=776\mev$: $\delta V_1$ is the $\bar{B} \Delta$ box, $\delta V_2$ is the $\bar{B} N$ box. (unit: MeV)}
\label{tab:polepCoupbox2}
\end{table}

We can see that the effect of the box with $\bar{B} \Delta$ or $\bar{B}^* \Delta$ intermediate states is a reduction of the mass of the lower state by bout $10-15\mev$, with an extra reduction of a about $15-25\mev$ from the box with intermediate $\bar{B} N$ and $\bar{B}^* N$ states. The upper state is not much modified by the box diagrams.

\section{Summary of the results}

Finally, since we have many intermediate results, we summarize here the final results that we get for the states, with $q_{max}=776\mev$, which we used to fix one of the $\Lambda_b$ energies. The results are shown in Table \ref{tab:polesum}, where we also write for a quick intuition the main channel of the state.
\begin{table}[H]
     \renewcommand{\arraystretch}{1.5}
     \setlength{\tabcolsep}{0.3cm}
\centering
\begin{tabular}{|c|c|c|c|c|}
\hline
main channel & $J$   &  $\quad I \quad$  &   $\, (E, \, \Gamma)\ \text{[MeV]} \,$  & Exp.   \\
\hline
$\bar{B} N$  &  1/2  & 0  & $5820.9, \, 0$ &  -  \\
\hline
$\pi \Sigma_b$  &  1/2  & 0  & $5969.5, \, 49.2$ & -   \\
\hline
$\bar{B}^* N$  &  1/2  & 0  & $5910.7, \, 0$ & $\Lambda_b(5912)$   \\
\hline
$\bar{B^*} N$  &  3/2  & 0  & $5920.7, \, 0$ &  $\Lambda_b(5920)$  \\
\hline
$\rho \Sigma_b$  &  1/2  & 0  & $6316.6, \, 2.8$ & -   \\
\hline
$\rho \Sigma_b$  &  3/2  & 0  & $6315.7, \, 3.8$ & -   \\
\hline
$\bar{B} N, \, \pi \Sigma_b$  &  1/2  & 1  & $6179.4, \, 122.8$ & -   \\
\hline
$\pi \Sigma_b$  &  1/2  & 1  & $6002.8, \, 132.4$ & -   \\
\hline
$\bar{B} \Delta, \, \pi \Sigma_b^*$  &  3/2  & 1  & $5932.9, \, 0$ & -   \\
\hline
$\pi \Sigma_b^*$  &  3/2  & 1  & $6063.8, \, 167.0$ & -   \\
\hline
$\bar{B^*} N$  &  1/2, 3/2  & 1  & $6202.2, \, 0$ & -   \\
\hline
$\rho \Sigma_b$  &  1/2, 3/2  & 1  & $6477.2, \, 10.0$ & -   \\
\hline
$\bar{B}^* \Delta$  &  1/2, 3/2, 5/2  & 1  & $6022.9, \, 0$ & -   \\
\hline
$\rho \Sigma_b^*$  &  1/2, 3/2, 5/2  & 1  & $6491.7, \, 1.6$ & -   \\
\hline
\end{tabular}
\caption{Energies and widths of the states obtained and the channels to which the states couple most strongly.}
\label{tab:polesum}
\end{table}

In summary, we predict 6 states with $I=0$, two of them corresponding to the $\Lambda_b(5912)$ and $\Lambda_b(5920)$, and 8 states with $I=1$. The energies of the states range from about 5800 MeV to 6500 MeV.

It is interesting to compare the results obtained here with those of Ref. \cite{juanloren}. In this later work, the same interaction as here is used for the main diagonal channels, but the transition between different coupled channels is not obtained through vector or pion exchange as done here, but invoking a combined SU(6) and heavy quark spin symmetry. In Ref. \cite{juanloren} the states of $I=1$ are not investigated but for $I=0$ four states are obtained, two of them, with $J=1/2,\ 3/2$, are also associated to the $\Lambda_b(5912)$ and $\Lambda_b(5920)$. In spite of the differences in the input, there are common features in the results. The two states associated to the $\Lambda_b(5912)$ and $\Lambda_b(5920)$ exhibit, as here, a substantial coupling to $\bar B^* N$. There is also a $1/2^-$ state in Ref. \cite{juanloren} at 5797 MeV which we find at 5820 MeV, only 33 MeV higher, and another state at 6009 MeV that we find at 5969 MeV, 40 MeV below. The mostly $\rho \Sigma_b$ state found here at 6316 MeV, basically degenerate in $J=1/2,\ 3/2$, was either not found or not searched for in Ref. \cite{juanloren} because of its higher mass. The qualitative agreement between the results of the two approaches is remarkable and gives further support to the common predictions. In addition, we have investigated states of $I=1$ and we find quite a few , some of them narrow enough for a clear experimental observation.

\section{Conclusions}
In this work we studied the interaction of $\bar B N$, $\bar B \Delta$, $\bar B^* N$ and $\bar B^* \Delta$ states with its coupled channels using dynamics mapped from the light quark sector to the heavy one. The starting point was to assume that the heavy quarks act as spectators in the dominant terms of the interaction. Then we studied pion exchange and vector exchange and obtained the results of pion exchange in the heavy quark spin symmetric approach combined with results of lattice QCD. The same procedure led us to the standard interaction of the extended local hidden gauge approach from the exchange of light vector mesons. With these elements of the interaction, supplemented with the subleading terms, in the large heavy quark mass counting, obtained from the exchange of heavy mesons in the local hidden gauge approach, we studied the interaction of the $\bar B N$, $\bar B \Delta$, $\bar B^* N$ and $\bar B^* \Delta$ with their coupled channels $\pi \Sigma_b$, $\pi \Lambda_b$, $\eta \Sigma_b$ (for the $\bar B N$); $\pi \Sigma_b^*$, $\eta \Sigma_b^*$ (for the $\bar B \Delta$); $\rho \Sigma_b$, $\omega \Lambda_b$, $\phi \Lambda_b$, $\rho \Sigma_b^*$, $\omega \Sigma_b^*$, $\phi \Sigma_b^*$ (for the $\bar B^* N$); and $\rho \Sigma_b^*$, $\omega \Sigma_b^*$, $\phi \Sigma_b^*$ (for the $\bar B^* \Delta$), and we searched for poles of the scattering matrix in different states of spin and isospin. We found six states in $I=0$, with one of them degenerate in spin $J=1/2,\ 3/2$, and eight states in $I=1$, less bound, two of them degenerate in spin $J=1/2,\ 3/2$, and two more degenerate in spin $J=1/2,\ 3/2,\ 5/2$. The coupling of the states to the different channels were evaluated, together with their wave function at the origin, in order to get a feeling of which are the largest building blocks in those molecular states. In particular, when studying the interaction of $\bar B^* N$ with its coupled channels, we found two states in $I=0$, degenerate in spin $J=1/2,\ 3/2$, which couple mostly to  $\bar B^* N$. The degeneracy is broken when the pion exchange is considered, allowing a mixture with intermediate $\bar B N$ states. Then we find that the mass of the states is close to 5910 MeV for natural values of the regularizing cutoff. More important, the splitting between the two states when the pion exchange is considered is found of the order of 10 MeV, rather independent of the cutoff used. This feature was considered relevant in view of the existence of the  $\Lambda_b(5912)$ and $\Lambda_b(5920)$ states, which are separated by this amount of energy. A fine tuning of the cut off was then done to match the exact energy of one of these states and then it was used subsequently to make predictions in all the other sectors of spin, isopin, leading to the states reported above.
We think that the use of realistic dynamics, with strict fulfilment of heavy quark spin-flavour symmetry, make the results obtained here rather accurate and should serve as a guideline for future experimental searches of baryon states with open beauty.

\section*{Acknowledgments}

We would like to thank J. Nieves for useful discussions, and Carmen Garcia-Recio, Laura Tolos, and Jia-Jun Wu for useful comments.
This work is partly supported by the Spanish Ministerio de Economia y Competitividad and European FEDER funds under Contract No. FIS2011-28853-C02-01 and the Generalitat Valenciana in the program Prometeo, 2009/090. We acknowledge the support of the European Community-Research Infrastructure Integrating Activity Study of Strongly Interacting Matter (Hadron Physics 3, Grant No. 283286) under the Seventh Framework Programme of the European Union.
This work is also partly supported by the National Natural Science Foundation of China under Grant No. 11165005 and by a scientific research fund (201203YB017) of the Education Department of Guangxi.

\end{document}